\documentclass[acmsmall]{acmart}

\usepackage{algorithmic}
\usepackage{algorithm}
\usepackage{textcomp}
\usepackage{stfloats}
\usepackage{url}
\usepackage{verbatim}
\usepackage{graphicx}
\usepackage{tabularx}
\usepackage{xcolor}
\usepackage{subcaption}

\usepackage{amsmath}
\usepackage{pifont}
\usepackage{multirow}
\usepackage{multicol}
\usepackage[normalem]{ulem}
\useunder{\uline}{\ul}{}
\usepackage{balance}
\usepackage{makecell}
\usepackage{color}
\usepackage[marginal]{footmisc}

\usepackage{varwidth}
\usepackage{caption}
\usepackage{graphicx}
\usepackage{float} 
\usepackage{marvosym}
\usepackage{bm}
\usepackage{enumitem}
\usepackage{colortbl} 
\usepackage{subcaption}
\usepackage[most]{tcolorbox}
\usepackage[table,xcdraw,dvipsnames]{xcolor}

\AtBeginDocument{%
  }

\setcopyright{acmlicensed}
\copyrightyear{2018}
\acmYear{2018}
\acmDOI{XXXXXXX.XXXXXXX}
\acmConference[Conference acronym 'XX]{Make sure to enter the correct
  conference title from your rights confirmation email}{June 03--05,
  2018}{Woodstock, NY}
\acmISBN{978-1-4503-XXXX-X/2018/06}

\begin{document}

\title{Ahead of the Spread: Agent-Driven Virtual Propagation for Early Fake News Detection}

\author{Bincheng Gu}
\email{gubincheng@stu.cqu.edu.cn}
\affiliation{%
  \institution{Chongqing University}
  \city{Chongqing}
  \country{China}
}

\author{Min Gao}
\email{gaomin@cqu.edu.cn}
\authornote{Corresponding author}
\affiliation{%
  \institution{Key Laboratory of Dependable Service Computing in Cyber Physical Society (Chongqing University), Ministry of Education}
  \city{Chongqing}
  \country{China}}

\author{Junliang Yu}
\affiliation{%
  \institution{The University of Queensland}
  \city{Brisbane}
  \country{Australia}}
\email{jl.yu@uq.edu.au}

\author{Zongwei Wang}
\email{zongwei@cqu.edu.cn}
\affiliation{%
  \institution{Chongqing University}
  \city{Chongqing}
  \country{China}
}

\author{Zhiyi Liu}
\email{liuzhiyi@stu.cqu.edu.cn}
\affiliation{%
  \institution{Chongqing University}
  \city{Chongqing}
  \country{China}
}

\author{Kai Shu}
\email{kai.shu@emory.edu}
\affiliation{%
  \institution{Emory Unversity}
  \city{Atlanta}
  \country{United States}
}

\author{Hongyu Zhang}
\email{hyzhang@cqu.edu.cn}
\affiliation{%
  \institution{Key Laboratory of Dependable Service Computing in Cyber Physical Society (Chongqing University), Ministry of Education}
  \city{Chongqing}
  \country{China}
}

\renewcommand{\shortauthors}{Gu, Gao, et al.}

\begin{abstract}
Early detection of fake news is critical for mitigating its rapid dissemination on social media, which can severely undermine public trust and social stability. Recent advancements show that incorporating propagation dynamics can significantly enhance detection performance compared to previous content-only approaches. However, this remains challenging at early stages due to the absence of observable propagation signals. To address this limitation, we propose AVOID, an \underline{a}gent-driven \underline{v}irtual pr\underline{o}pagat\underline{i}on for early fake news \underline{d}etection. AVOID reformulates early detection as a new paradigm of evidence generation, where propagation signals are actively simulated rather than passively observed. Leveraging LLM-powered agents with differentiated roles and data-driven personas, AVOID realistically constructs early-stage diffusion behaviors without requiring real propagation data. The resulting virtual trajectories provide complementary social evidence that enriches content-based detection, while a denoising-guided fusion strategy aligns simulated propagation with content semantics. Extensive experiments on benchmark datasets demonstrate that AVOID consistently outperforms state-of-the-art baselines, highlighting the effectiveness and practical value of virtual propagation augmentation for early fake news detection. The code and data are available at \url{https://github.com/Ironychen/AVOID}.
\end{abstract}

\ccsdesc[500]{Information systems~Data mining}
\ccsdesc[500]{Computing methodologies~Natural language processing}

\keywords{Fake News Detection, Social Simulation,
LLM-Based Agents}


\maketitle

\section{Introduction}
Social media platforms such as Facebook and X have become the primary channels for news consumption, enabling rapid and large-scale information dissemination. Although this increases accessibility, it also accelerates the spread of fake news, erodes public trust, and distorts social perceptions \cite{fisher, shu2017fake}. The rapid dissemination of misinformation, particularly in sensitive areas such as politics and public health \cite{allcott2017social, roozenbeek2020susceptibility}, can have serious societal consequences before corrective interventions can be implemented \cite{vosoughi2018spread}. These challenges highlight the urgent need for effective methods that can detect fake news at the earliest possible stage. 

\begin{figure}[t]
    \centering
    \includegraphics[width=\linewidth]{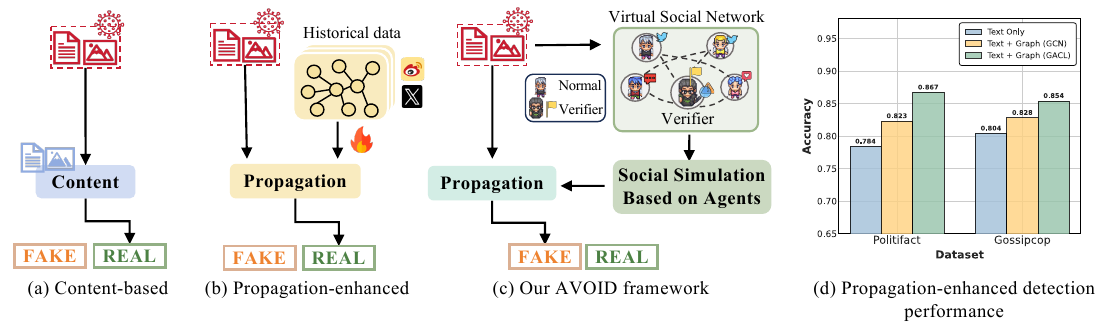}
    \caption{Figures (a)-(c) compare the architectures of different fake news detection methodologies, while Figure (d) presents the comparative performance across two datasets, confirming the significant enhancement provided by propagation-enhanced methods over content-only baselines.}
    \Description{A bar chart showing that graph-based models outperform content-only models, and user information leads to further improvement.}
    \label{fig_intro1}
\end{figure}

Mainstream methods for early fake news detection have primarily relied on content-based analysis \cite{textcnn, zhang2021mining, cafe}. As shown in Figure~\ref{fig_intro1}(a), these approaches utilize textual, visual, and emotional features extracted directly from the news content to assess credibility. However, purely content-based methods have inherent shortcomings: They often fail to capture the broader contextual and social dynamics involved in news dissemination, which reduces their ability to distinguish sophisticated fake news from genuine information, especially when deceptive news closely imitates credible writing styles~\cite{zhang2020adversarial}. To mitigate the lack of contextual cues in content-only methods, recent work has explored the use of social propagation patterns as complementary signals for detection~\cite{bian2020rumor, sun2022rumor}. Extending this direction to early-detection scenarios where real-time data is scarce, existing studies utilize deep graph generative models to synthesize virtual propagation trajectories \cite{you2018graphrnn,zhang2019d,zhang2024mitigating}, as depicted in Figure~\ref{fig_intro1}(b), which serve as auxiliary information to enrich the representation of news articles during detection. Our empirical results, shown in Figure~\ref{fig_intro1}(d), directly compare the performance of content-only models and those integrating propagation information, confirming that the inclusion of propagation signals significantly improves detection accuracy.


Despite the promise of generative approaches, they face a critical limitation: most existing graph generative models focus on fitting the statistical distributions of historical graph structures, often failing to capture the dynamic, content-driven interactions between users and news articles. Consequently, they struggle to produce reliable, trustworthy trajectories for entirely new articles with novel contexts~\cite{you2018graphrnn, zhang2024mitigating}.  In practice, the ideal scenario is to detect fake news before any real dissemination occurs, a situation in which content-based methods become the only feasible alternative. To bridge this gap, our study proposes a novel paradigm to generate the necessary evidence for a definitive judgment on difficult, low-confidence news articles by leveraging agent-based simulation to construct virtual propagation trajectories through modeling realistic user-content interactions within a social network environment.

Recent advances in Large Language Models (LLMs) and agent-based modeling techniques \cite{achiam2023gpt, llama, li2024agent} have enabled more realistic simulation of complex social interactions and information propagation processes. Emerging research has shown that agent-driven simulations can effectively emulate individual user behaviors and specific social phenomena observed in real-world environments, providing a powerful tool for studying social dynamics \cite{chen2024large, park2023generative,ju2025trajllm, wang2025yulan}. Built on these advancements, we attempt to simulate user behaviors and interactions from scratch, creating plausible virtual propagation paths even in the absence of observed data. Despite the progress in social simulation, substantial obstacles to achieving realistic and reliable virtual propagation remain. Current agent-based simulation approaches often initialize agents using basic sociodemographic profiles and assign them uniform behavioral rules \cite{liu2024skepticism, mou2024unveiling}. Such agent initialization methods based solely on statistical features overlook the nuanced complexity of real-world user behaviors in social interactions. Furthermore, the simulated propagation trajectory generation process can introduce noise, which undermines the effectiveness of early detection.

To overcome these limitations, we propose AVOID, a LLM-Empowered \underline{a}gent-driven \underline{v}irtual pr\underline{o}paga-t\underline{i}on for early and robust fake news \underline{d}etection framework without relying on any observed diffusion data, as illustrated in Figure~\ref{fig_intro1}(c). We categorize participating agents into two distinct types: \textit{Diffuser Agents}, representing the general user population who drive the information spread through passive browsing and forwarding, and \textit{Verifier Agents}, who possess the capability to critically assess content veracity and actively intervene. By explicitly distinguishing and modeling these roles, AVOID faithfully simulates the interaction heterogeneity commonly observed in real social platforms. In addition, as suggested by~\cite{liu2025mosaic, kiesling2012agent, zhang2019empirically, shi2025personax}, user personas are a key determinant of interaction behavior, and aligning persona distributions with empirical observations is crucial for achieving realistic simulated behaviors and diffusion dynamics. Accordingly, we align agent personas with real user data. Furthermore, to alleviate noises introduced by virtual simulation, AVOID applies a denoising-guided cross-modal fusion strategy, thereby enhancing the reliability and accuracy of early fake news detection. 

Overall, our contributions can be summarized as follows:
\begin{itemize}
    \item We present a novel early fake news detection paradigm that incorporates virtual propagation augmentation even without any observed diffusion paths, addressing significant limitations of existing approaches.
    \item We propose AVOID, a LLM-driven agent framework that enhances early fake news detection by generating agent-driven virtual propagation networks derived from realistic personas and mitigating the noise inherent in these paths through denoising-guided cross-modal fusion.
    \item Extensive experiments conducted on real-world datasets confirm that AVOID consistently outperforms state-of-the-art baselines, validating the effectiveness and practicality of our virtual propagation-based approach for early fake news detection.
\end{itemize}

This article is structured as follows: Section ~\ref{sec_Preliminaries} introduces the preliminaries, including the agent-based social network setting, propagation-enhanced fake news detection, and the confidence-based filtering strategy. Section ~\ref{sec_Agentrole} describes the design of diffuser and verifier agents. Section ~\ref{sec_AVOID} presents the proposed AVOID framework. Section ~\ref{sec_experiments} reports the experimental setup and results. Section ~\ref{sec_related} reviews related work, and Section ~\ref{sec_conclusion} concludes the paper and outlines future directions.

\section{Preliminaries}
\label{sec_Preliminaries}

In this section, we first outline the agent-based social network simulation setting, then formalize propagation-enhanced fake news detection, and finally define hard news together with a filtering strategy that routes only low-confidence items to generate virtual propagation, ensuring computational efficiency.

\subsection{Agent-Based Social Network Simulation}
Leveraging the generative and reasoning abilities of LLMs, autonomous agents simulate human behavior in complex social contexts~\cite{aher2023using, park2023generative, xie2024can}. In our study, we adopt a typical agent architecture that consists of a profile module, a memory module, and an action module \cite{wang2024survey}. The profile module defines the identity and behavior of the agent, the memory module captures interaction history, and the action module translates reasoning into observable behaviors like commenting or reposting. 

We model the social environment as a network of interacting agents $\mathcal{V}$. 
Each node $v_i \in \mathcal{V}$ corresponds to an autonomous agent characterized by a unique profile, memory, and action strategy. News spread through interactions among neighboring agents. Specifically, when an agent $v_i$ shares a news item and successfully influences a neighbor $v_j \in \mathcal{N}(v_i)$ to engage with the content, a directed edge $(v_i \rightarrow v_j)$ is established. For each news item $i$, the resulting interactions induce a dynamic propagation graph $\mathcal{G}_i = (\mathcal{V}_i, \mathcal{E}_i)$, where $\mathcal{V}_i \subseteq \mathcal{V}$ denotes the set of participating agents and $\mathcal{E}_i$ comprises the corresponding influence edges. Each propagation graph $\mathcal{G}_i$ captures the temporal diffusion trajectory of the news item.

\subsection{Propagation Enhanced Fake News Detection} 
\label{sec2.2}
Fake news detection has traditionally relied on content-based features derived from multimodal information \(\mathcal{X_C}\) present in news content. These methods typically employ encoders to obtain semantic representations of \(\mathcal{X_C}\), which are then used to learn the conditional probability distribution \(P_{\theta}(y \mid \mathcal{X_C})\) to predict the veracity of the news.

To further enhance detection capabilities, recent studies \cite{dou2021user, zhu2024propagation} have incorporated propagation dynamics from social networks. By modeling the diffusion patterns of news as structural features derived from propagation graphs \(\mathcal{G}\), these propagation-enhanced methods extend the input space for detection. Consequently, the detection objective expands to \(P_{\theta}(y \mid \mathcal{X_C}, \mathcal{X_G})\), where \(\mathcal{X_G}\) captures the topological characteristics of its diffusion within the social network.

Both content-centric and propagation-enhanced approaches share a common training paradigm: minimizing the binary cross-entropy loss. Formally, the detection model optimizes:
\begin{equation}
\mathcal{L}_{cls} = -\sum_{i=1}^{Z} \left[y_i \log(\hat{y}_i) + (1 - y_i) \log(1 - \hat{y}_i)\right],
\label{cls}
\end{equation}
where \( \hat{y}_i \in \hat{Y} \) is the predicted probability that the \( i \) -th news article is fake, \( y_i \in Y \) is the corresponding ground-truth label, and \( Z \) is the total number of news samples.

\subsection{Confidence-Guided Selective Virtual Propagation} 
\label{filtering}
Although virtual propagation can provide valuable additional evidence for fake news detection, it is computationally expensive. To avoid simulating propagation on easily decidable items, we introduce a filtering strategy that detects ``hard'' news cases. Concretely, we start from a pretrained BERT encoder and fine-tune only its last two transformer layers on the target training dataset to obtain a lightweight content classifier $f_{\theta}$. Given an input news item $x$, the classifier produces a posterior probability $\hat{p}(x)=P_{\theta}(y \mid \mathcal{X}_C)$, where $\mathcal{X}_C$ denotes the text representation encoded by BERT. We define the confidence as $conf(x) = \max\{\hat{p}(x), 1-\hat{p}(x)$, which measures how strongly the classifier favors either class. If $conf(x)$ is below the threshold, the sample is regarded as a low-confidence (hard) instance. Only for these instances do we invoke the virtual propagation module to generate the graph-based representation $\mathcal{X}_G$. The final prediction is then made using the joint representation $(\mathcal{X}_C, \mathcal{X}_G)$ as described in Section~\ref{sec2.2}. For high-confidence (easy) samples, we skip propagation entirely and rely solely on $f_{\theta}$, thereby achieving a favorable balance between accuracy and efficiency.

\begin{table}[tbp]
\centering
\caption{Comment filtering statistics. We use a rule-based filter to separate low-information reactions from informative comments that contain interpretation or verification cues. Filtered reports the number of retained informative comments, and the percentage (\%) is the retained ratio over all comments.}
\begin{tabular}{lccc}
\toprule
Dataset & Total Comments & Filtered &Percentage (\%) \\
\midrule
PolitiFact  & 229{,}370  & 8{,}778  & 4.83 \\
GossipCop   & 180{,}902  & 4{,}098  & 2.26 \\
Weibo  & 367{,}070  & 20{,}475  & 5.58 \\
\bottomrule
\end{tabular}
\label{tab:comment_stats}
\end{table}

\section{Agents for Propagation Simulation}
\label{sec_Agentrole}

To simulate realistic information propagation, we introduce two agent roles that reflect heterogeneous participation in social networks. Previous studies \cite{bakshy2012role} and our analysis (Table~\ref{tab:comment_stats}) show that the vast majority of responses are low-information reactions, whereas only a small fraction contain meaningful cues that can steer downstream discussion. Building on this observation, we design two types of agents: diffuser agents emulate lightweight, reaction-driven participants, while verifier agents produce more informative responses with enhanced truth assessment and subjective expression. This role-based design enables more faithful modeling of heterogeneous user behaviors in rumor propagation. 

\subsection{Design of Diffuser Agent}
    Diffuser agents represent the majority of users on social networks, engaging with information based on their experiences, interests, and beliefs. In our framework, they form the basis of the simulated network to reflect diverse user behaviors.

\subsubsection{Diverse User Profile Modeling} 

Most existing simulations rely on randomly initialized agent profiles derived from statistical features, which do not reflect the true persona distribution for specific types of news \cite{park2023generative, mou2024unveiling}. To overcome this limitation, we extract fine-grained persona representations from real-world datasets to better align with the actual distribution of user characteristics. The detailed methodology is presented in Section~\ref{persona}. These fine-grained personas are further augmented with social context metadata, enhancing realism and enabling nuanced simulations.

\subsubsection{Temporal Contextual Memory Modeling.}

Human social behavior depends not only on the current context but also on past experiences. To capture this temporal dependency, we introduce short-term and long-term memories to simulate social exposure.

\textit{Simulating Short-Term Social Interaction.} To reflect recent social interactions, this module records the latest \(k\) social interactions, such as news summaries and comments of friends, as vector embeddings. When encountering a new item, it uses FAISS \cite{douze2024faiss} to efficiently retrieve the most similar records based on cosine similarity, providing immediate context for decision-making.

\textit{Simulating Long-Term Knowledge Accumulation.} Long-term memory maintains distilled knowledge by periodically summarizing short-term content into compact records, retaining essential experiences while a forgetting mechanism discards outdated information. For retrieval, it uses an LLM to decompose new items into semantic sub-queries (entities, events, topics). This enables more precise, context-aware retrieval of relevant past knowledge, unlike the direct vector similarity search used in short-term memory. The guiding prompt is shown below: 
\begin{center}
\begin{tcolorbox}[
  enhanced,            
  colback=gray!10,
  colframe=gray,
  width=0.98\linewidth,
  arc=1mm,
  auto outer arc,
  title={Long-term Memory Retrieval Prompt Template},
  breakable,
  boxsep=0.2mm           
]
Given the \textcolor{RoyalBlue}{$<$News Content$>$}, decompose it into \textcolor{PineGreen}{$<$Entity$>$}, \textcolor{PineGreen}{$<$Event$>$}, and \textcolor{PineGreen}{$<$Topic$>$}, then retrieve relevant segments from \textcolor{purple}{$<$Long-Term Memory$>$}.
\end{tcolorbox}
\end{center}

\subsubsection{User Engagement Action Modeling.} 

Simulating user behavior on social networks requires agents to actively engage with news content. To this end, we design an action module that supports the following behaviors. For diffuser agents, available actions include (1) \textit{comment}: writing a textual response to express the stance and provide a brief explanation; (2) \textit{forward}: sharing an existing post, either directly or with an added remark; (3) \textit{like}: expressing approval by liking a post; and (4) \textit{view}: only viewing the news without taking any further action. These actions influence the agent's exposure and indirectly shape the diffusion process.

\subsection{Design of Verifier Agent}
While diffuser agents represent the general user population, verifier agents model a small subset of influential users who critically shape the information flow and opinion dynamics, capturing their proactive and strategic influence on news dissemination.

\subsubsection{Specific Verifier Profile Modeling}
\label{verifier_profile}

Unlike diffuser agents, verifier personas are derived explicitly from influential users. We first compute an influence score based on repost counts, average likes, and follower numbers, and label the top 5\% of users as influential. From these users' historical comments, we keep only those that have strong engagement signals and contain clear, well-supported stances, such as comments that cite official sources, scientific publications, or government statistics. We then apply the method in Section~\ref{persona} to the filtered comments to extract verifier-specific persona profiles.

\subsubsection{Enhanced Policy Memory Modeling.}
To enable advanced reasoning for credibility assessment, verifier agents maintain a policy memory that stores structured traces of past veracity judgments, including content, stances, rationales, and social context. Memory is hierarchically organized into three levels: entity level \( \pi_e \) to track key actors, event level \( \pi_v \) to capture causal and temporal patterns, and meta-level \( \pi_m \) to summarize general reasoning strategies. This design improves the ability of verifiers to make a judgment based on context and have a credible influence within the network.

\subsubsection{Veracity-Driven Action Modeling}

In addition to the basic actions available to all users, verifier agents are endowed with two additional actions that reflect their authoritative role in real-world networks: (5) \textit{fact-check}: verifying news items using external knowledge sources or fact-checking tools; and (6) \textit{warning}: issuing public warnings to suppress the spread of misinformation. These privileged actions enable verifier agents to actively intervene and reshape the trajectory of information propagation.

\section{AVOID: Agent-Driven News Virtual Propagation}
\label{sec_AVOID}


This section introduces AVOID, a simulation framework designed for early fake news detection by generating and utilizing virtual propagation trajectories. Note that virtual propagation is only conducted on "low-confidence" samples introduced in Section \ref{filtering} to balance detection effectiveness and computational efficiency.

\subsection{Realistic Persona Extraction}
\label{persona}
To initialize diffuser and verifier agents with realistic user behaviors, we extract agent personas from real-world social media comments. We adopt a multi-stage persona extraction pipeline that progressively organizes, samples, and distills user comments into coherent and behaviorally grounded profiles. To ensure a leakage-free process, persona extraction and agent initialization are restricted to the training data and performed without news veracity labels.

\subsubsection{Persona-Oriented Hierarchical Clustering}

To distill representative user archetypes from the raw news-comment content, we propose a two-stage hierarchical clustering process that transitions from thematic discovery to viewpoint extraction. 

The process begins with context-level grouping, where the news content is embedded into a semantic space and partitioned into $m$ topic clusters $\mathcal{T}=\{t_1,\dots,t_m\}$. Building upon these contexts, we perform viewpoint-level distillation within each cluster. For a topic $t_k$, let $\mathcal{R}_k=\{r_{k,1},\dots,r_{k, n_k}\}$ denote its associated user comments. We cluster the comment embeddings to capture distinct reaction patterns, and map each cluster back to its member comments, producing a set of comment groups $\mathcal{P}_k=\{p_{k,1},\dots,p_{k,\ell_k}\}$. Here, each group $p_{k,j}\subseteq\mathcal{R}_k$ aggregates comments with similar viewpoints, corresponding to a topic-specific persona. The global persona pool is defined as $\mathcal{P}=\bigcup_{k=1}^{m}\mathcal{P}_k$. This hierarchical approach ensures that the extracted personas are diverse in perspective and grounded in the specific news topics they address.

\subsubsection{Balanced Persona Sampling}

Although hierarchical clustering yields diverse persona groups, using all comments within a specific group $p_{k,j}$ can introduce redundancy. Centroid-based selection~\cite{sorscher2022beyond} can oversimplify user profiles, while boundary-focused methods~\cite{paul2021deep} risk overfitting to outliers. To mitigate these issues, we adopt a Balanced Persona Sampling strategy. For each persona group $p_{k,j}$, we denote its constituent comment embeddings as $\mathbf{E}_{k,j} = \{ \mathbf{e}_i \mid r_i \in p_{k,j} \}$. We then select a compact yet representative subset of comments $\mathcal{R}^*_{k,j} \subseteq p_{k,j}$ by optimizing for both prototypicality and diversity. This leads to the following optimization problem:
\begin{equation}
\max_{\mathcal{R}^*_{k,j}} \left(
w_p \!\!\sum_{r_i \in \mathcal{R}^*_{k,j}} \!\frac{1}{1{+}D(\mathbf{e}_i,\boldsymbol{\mu}_{k,j})}
+ w_d \cdot \frac{2}{|\mathcal{R}^*_{k,j}|} \!\!\sum_{r_a, r_b \in \mathcal{R}^*_{k,j}} \! D(\mathbf{e}_a,\mathbf{e}_b)
\right),
\end{equation}
where $\boldsymbol{\mu}_{k,j}$ is the centroid of $\mathbf{E}_{k,j}$ in the embedding space. The term $D$ represents the Euclidean distance, while weights $w_p$ and $w_d = 1 - w_p$ balance prototypicality and diversity. The selected subset $\mathcal{R}^*_{k,j}$ captures the core semantics and variations of the underlying persona. We solve this discrete optimization problem using a greedy algorithm that iteratively selects candidates with the maximum marginal gain. While not guaranteeing global optimality, this approach consistently yields a compact representative subset in our experiments.

\begin{figure*}[t]
    \centering
    \includegraphics[width=\linewidth]{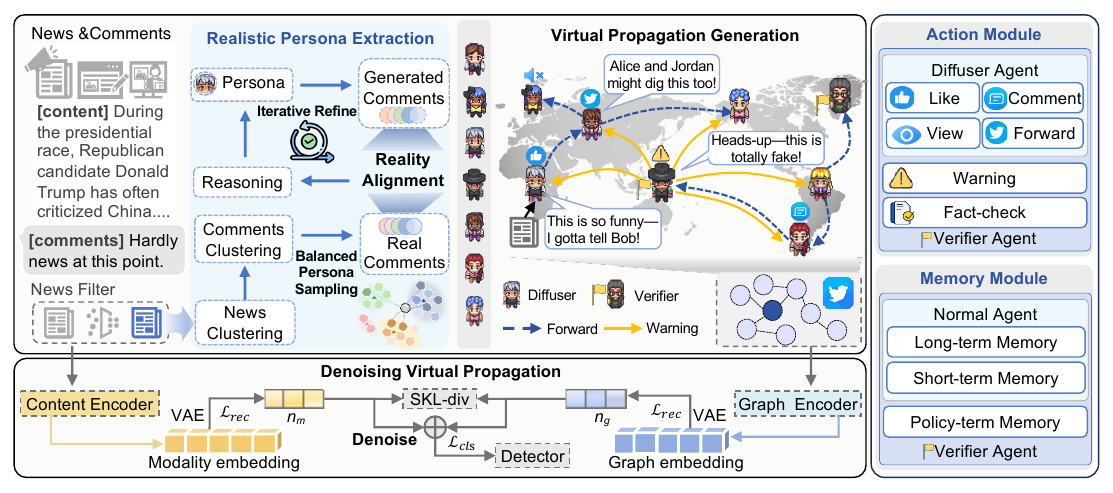}
    \caption{Overview of AVOID. Real-world comments are distilled into fine-grained personas, which initialize simulated agents. These agents participate in virtual propagation to facilitate early fake news detection.}
    \label{framework}
    \Description{framework.}
\end{figure*} 

\subsubsection{Reality-Aligned Persona Distilling.}
From each sampled comment subset $\mathcal{R}^*_{k,j}$, we derive a fine-grained persona profile for the corresponding user cluster. Since directly prompting LLMs with raw comments often results in generic or superficial personas, we employ an iterative, LLM-guided reflection and refinement procedure to improve social realism and behavioral plausibility. At each iteration $t$, we sample a small batch of comments from $\mathcal{R}^*_{k,j}$ as evidence to progressively refine the persona profile:
\begin{equation}
\mathcal{P}_{t+1} \leftarrow \text{Reflection}(\mathcal{P}_{t}, \hat{r}_t, \boldsymbol{\mu}_t, \lambda),
\end{equation}
where $\hat{r}_t$ is the comment generated by the LLM conditioned on the current profile $\pi_t$, and $\boldsymbol{\mu}_t$ denotes the semantic center of the sample batch. The reflection mechanism is triggered when the cosine similarity between the generated comment embedding $\mathrm{Embed}(\hat{r}_t)$ and the center $\boldsymbol{\mu}_t$ falls below a threshold $\lambda=0.7$. This discrepancy prompts the LLM to revise $\mathcal{P}_{t}$ to better align with the observed evidence. The system prompt used is as follows:

\begin{center}
\begin{tcolorbox}[
  enhanced,
  colback=gray!10,
  colframe=gray,
  width=0.98\linewidth, 
  arc=1mm,
  auto outer arc,
  title={Person Distillation Prompt Template},
  breakable,
  boxsep=0mm,
  top=2mm,
  bottom=2mm,
  fontupper=\linespread{0.9}\selectfont
]
Given the current \textcolor{RoyalBlue}{$<$Persona$>$}, you generated a representative \textcolor{PineGreen}{$<$Comment$>$} that conflicts with the \textcolor{PineGreen}{$<$Next Comment$>$}. Reflect on the \textcolor{purple}{$<$Reason$>$} and method to refine the persona, and output the updated \textcolor{RoyalBlue}{$<$Persona$>$}.
\end{tcolorbox}
\end{center}
Through this process, we extract realistic fine-grained persona representations from real-world datasets, aligning the simulated agent distribution with that observed in actual social environments. Note that the persona extraction is a one-time offline process and incurs negligible computational overhead.

\subsection{Virtual Propagation Generation}

Building on the extracted personas assigned to each agent, this section details the simulation of news propagation within the social network, resulting in structured diffusion graph data for downstream analysis.

\subsubsection{Initialization with Realistic Persona}
We construct the social graph by extracting a closed subgraph from a real-world network to preserve realistic user relationships. Each agent is initialized with a profile of the sampled persona and social attributes. Initially, all agents have empty short-term and long-term memories. Verifier agents are additionally assigned policy memory and access to fact-checking tools for veracity assessment. 

\subsubsection{Social Dynamic Modeling with Memory Updating}
Memory updates enable agents to adapt by integrating social feedback and past experiences; for verifier agents, this process is further extended to include reflective self-correction based on judgment errors. In our social simulation, each agent’s short-term memory continuously records its friends’ real-time actions and comments on a news item. This mechanism provides the agent with an immediate and evolving snapshot of how news is perceived within its circle. To further model the long-term evolution of social hot topics, the agent periodically consolidates short-term records into a long-term knowledge base. During this process, significant events and interactions are distilled into higher-level insights representing current trends. When integrating these insights into long-term memory, we apply an exponential time-decay scheme that down-weights older information and prioritizes recent evidence, enabling continual adaptation to evolving social dynamics.

Leveraging the above memory dynamics, verifier agents further adapt through a policy update process that is triggered by incorrect veracity judgments. When a judgment is flawed, the agent employs Chain-of-Thought (CoT) reasoning \cite{wei2022chain} to rethink its original reasoning trace $\Phi_t$. By comparing this trace against the ground truth, the agent identifies the precise logical error in its initial assessment, which considered the news content $x_t$, its policy memory ${\Pi}_t$, and the social context $K_t$. This targeted error analysis directly informs the refinement of its policy memory to improve future strategies. Formally, the update is defined as:  
\begin{equation}
    \Pi_{t+1} \leftarrow \text{Reflection}(x_t, {\Pi}_t, K_t, \Phi_t),
\end{equation}
the Reflection function refines the policy's multilevel components (entity-level, event-level, and meta-level). When an error is detected, this process is operationalized through a specific prompt:
\begin{center}
    \begin{tcolorbox}[
  enhanced,           
  colback=gray!10,
  colframe=gray,
  width=0.98\linewidth,
  arc=1mm,
  auto outer arc,
  title={Policy Memory Refinement Prompt},
  breakable,
  boxsep=0mm,
  top=2mm,    
  bottom=2mm, 
  fontupper=\linespread
]
Given the \textcolor{RoyalBlue}{$<$News Content$>$}, \textcolor{RoyalBlue}{$<$Policy Memory$>$}, \textcolor{RoyalBlue}{$<$Social Context$>$}, and \textcolor{RoyalBlue}{$<$Ground Truth$>$}, first generate your assessment and \textcolor{PineGreen}{$<$Reason$>$}. If the assessment does not match \textcolor{RoyalBlue}{$<$Ground Truth$>$}, use Chain-of-Thought reasoning to analyze and then refine your \textcolor{purple}{$<$Policy$>$} (\textcolor{purple}{$\pi_e$}, \textcolor{purple}{$\pi_v$}, \textcolor{purple}{$\pi_m$}) based on the analysis.
\end{tcolorbox}
\end{center}
This reflective update enables verifier agents to dynamically evolve their decision-making strategies, leading to more accurate misinformation detection and greater influence on social networks.

\subsubsection{Action-Driven Propagation}
After memory initialization, agents propagate news by making stepwise decisions informed by memory and social context, thereby shaping the diffusion trajectory. For each news item, propagation begins from selected seed agents and unfolds in discrete steps. At each step, agents choose an action from a predefined set, guided by short- and long-term memories. The prompt used for decision-making is shown below:
\begin{center}
\begin{tcolorbox}[
  enhanced,           
  colback=gray!10,
  colframe=gray,
  width=0.98\linewidth,
  arc=1mm,
  auto outer arc,
  title={Diffuser Agent Action Prompt Template},
  breakable,
  boxsep=0mm,
  top=2mm,    
  bottom=2mm, 
  fontupper=\linespread
]
You are \textcolor{RoyalBlue}{$<$Persona$>$}. You just saw \textcolor{RoyalBlue}{$<$News Content$>$} shared by your friends. Considering your short-term memory \textcolor{purple}{$<$$\mathcal{M}^{S}$$>$} and Long-term memory \textcolor{purple}{$<$$\mathcal{M}^{L}$$>$}, select the most appropriate action from the \textcolor{PineGreen}{$<$Action List$>$}.
\end{tcolorbox}
\end{center}

In addition to the general decision-making mechanism, designated verifier agents follow specialized procedures: they assess forwarded news using learned policies and fact-checking tools. If misinformation is detected, they broadcast warnings to the network. The specific prompt is below:
\begin{center}
\begin{tcolorbox}[
  enhanced,            
  colback=gray!10,
  colframe=gray,
  width=0.98\linewidth,
  arc=1mm,
  auto outer arc,
  title={Verifier Agent Action Prompt Template},
  breakable,
  boxsep=0mm,
  top=2mm,    
  bottom=2mm, 
  fontupper=\linespread{0.9}\selectfont
]
You are \textcolor{RoyalBlue}{$<$Verified Agent Persona$>$} on a social network, responsible for veracity assessment. Use your \textcolor{PineGreen}{$<$Policies$>$} and \textcolor{PineGreen}{$<$Fact-Checking$>$} tool to verify the \textcolor{RoyalBlue}{$<$News Content$>$}. Then select an action from the \textcolor{PineGreen}{$<$Action List$>$}. \\
Action: \textcolor{purple}{WARN} — This news might be fake.
\end{tcolorbox}
\end{center}

Agents propagate news based on dynamically updated memories, generating a graph $\mathcal{G}_i$ for each news by simulating its spread from random seed agents within a fixed depth.

\subsection{Denoising Virtual Propagation For Detection}

With the virtual propagation paths generated for hard news, the next step is to effectively utilize this augmented evidence for veracity prediction. We first encode features from both the news content and the simulated graphs. Then, to maximize the utility of the simulated signals and bridge the discrepancy between content and graph modalities, we apply symmetric KL divergence for feature alignment. The theoretical validity of this approach is formally proved in Section~\ref{sec:theory_prove}.

\subsubsection{Encoding Content and Simulated Propagation} To capture the characteristics of each news item, we extract two types of features: a content representation $\mathcal{X}_C$ from the article text and a propagation representation $\mathcal{X}_G$ from the virtual interaction graph.

\textit{News Content Feature Extraction.}
We adopt a Hierarchical Attention Network \cite{han} to encode the semantic features of each news article. The article is partitioned into $S$ sentences, each containing up to $L$ words. We first utilize a pre-trained BERT encoder to generate contextualized word embeddings, which are then processed by a GRU to capture sequential dependencies. To derive the sentence representations, we employ a word-level attention mechanism:
\begin{equation}
\mathbf{s}_j = \sum_{t=1}^{L} \gamma_t^{(j)} \mathbf{h}_t^{(j)}, \quad
\gamma_t^{(j)} = \frac{\exp(\mathbf{h}_t^{(j)\top} \mathbf{u}_w)}{\sum_{t'=1}^{L} \exp(\mathbf{h}_{t'}^{(j)\top} \mathbf{u}_w)},
\end{equation}
where $\mathbf{h}_t^{(j)}$ is the GRU hidden state at time $t$ for sentence $j$, and $\mathbf{u}_w$ is a trainable word-level attention vector. The resulting sentence vectors $\mathbf{s}_j$ are subsequently fed into another GRU and a sentence-level attention mechanism to aggregate the final document-level representation $\mathbf{x}_c$:
\begin{equation}
\mathbf{x}_c = \sum_{j=1}^{S} \delta_j \mathbf{h}_j, \quad
\delta_j = \frac{\exp(\mathbf{h}_j^{\top} \mathbf{u}_s)}{\sum_{j'=1}^{S} \exp(\mathbf{h}_{j'}^{\top} \mathbf{u}_s)},
\end{equation}
where $\mathbf{h}_j$ denotes the sentence-level hidden state and $\mathbf{u}_s$ is a trainable sentence-level attention vector. This hierarchical architecture ensures that $\mathbf{x}_c$ captures both local word context and global document-level semantics.

\textit{Propagation Feature Extraction.}
We construct a virtual propagation graph with adjacency matrix $\mathbf{A}_s$ and node embeddings $\mathbf{v}_i$ initialized by BERT-encoded agent profiles. To capture structural dependencies, we employ Graph Attention Networks~\cite{velivckovic2017graph}, updating node embeddings at layer $l$ via:
\begin{equation}
\mathbf{v}_i^{(l+1)} = \sigma\!\left( \sum\nolimits_{j\in\mathcal{N}(i)} \alpha_{ij}^{(l)} \mathbf{U}^{(l)} \mathbf{v}_j^{(l)} \right),
\end{equation}
where $\mathbf{U}^{(l)}$ is the trainable weight matrix, $\sigma(\cdot)$ is the activation function, and $\mathcal{N}(i)$ denotes the neighbor set. $\alpha_{ij}^{(l)}$ represents the attention coefficient following the standard GAT formulation. Finally, we apply graph-level aggregation to obtain the propagation-aware representation $\mathbf{x}_g$.

\subsubsection{Denoising-Guided Alignment for Detection}
\label{denoise}

To mitigate artifacts in simulated virtual propagation, we align the multi-modal latent representations via symmetric KL divergence, treating the news content as a semantic anchor. This alignment serves as a denoising regularizer that suppresses modality-specific noise while distilling shared, veracity-related semantics.

Let $\mathbf{x}_c$ denote the content representation and $\mathbf{x}_g$ denote the (virtual) propagation representation. Since they are produced by different encoders and may lie in heterogeneous feature spaces, we first map them into a unified $d$-dimensional space:
\begin{equation}
\tilde{\mathbf{x}}_c = f_c(\mathbf{x}_c), \qquad
\tilde{\mathbf{x}}_g = f_g(\mathbf{x}_g),
\end{equation}
where $f_c(\cdot)$ and $f_g(\cdot)$ are linear projection layers. This step ensures that subsequent alignment and fusion are performed on comparable representations.

We then employ two Variational Autoencoders (VAEs) \cite{kingma2013auto} to map these aligned features into a latent space. Specifically, we model each modality as a combination of a shared semantic signal $\mathbf{s}$ and modality-specific noise $\bm{\epsilon}$:
\begin{equation}
    \tilde{\mathbf{x}}_c = \mathbf{s} + \bm{\epsilon}_c, \quad \tilde{\mathbf{x}}_g = \mathbf{s} + \bm{\epsilon}_g,
\end{equation}
where $\bm{\epsilon}_c, \bm{\epsilon}_g$ represent noise terms that are unbiased on average and uncorrelated with $\mathbf{s}$. Each modality is encoded into a Gaussian posterior over a latent variable $\mathbf{z}$:
\begin{equation}
    q_\phi(\mathbf{z} \mid \tilde{\mathbf{x}}_c) = \mathcal{N}\!\bigl(\bm{\mu}_c, \bm{\Sigma}_c\bigr), \quad
    q_\phi(\mathbf{z} \mid \tilde{\mathbf{x}}_g) = \mathcal{N}\!\bigl(\bm{\mu}_g, \bm{\Sigma}_g\bigr),
\end{equation}
where $\bm{\mu}_c,\bm{\mu}_g$ and $\bm{\Sigma}_c,\bm{\Sigma}_g$ denote the predicted mean vectors and covariance matrices of the modality-specific latent posteriors, respectively. To ensure that the latent variable $\mathbf{z}$ preserves the essential modality-specific semantics of $\tilde{\mathbf{x}}_c$ and $\tilde{\mathbf{x}}_g$, we introduce a reconstruction loss for each modality:
\begin{equation}
    \mathcal{L}_{\text{rec}}(\tilde{\mathbf{x}}) = \mathbb{E}_{q_\phi(\mathbf{z}\mid\tilde{\mathbf{x}})}\!\left[-\log p_\psi(\tilde{\mathbf{x}} \mid \mathbf{z})\right].
\label{rec}
\end{equation}
This objective encourages the latent variable $\mathbf{z}$ to retain essential information from the input $\tilde{\mathbf{x}}$ for reconstruction. Moreover, this formulation implicitly establishes a connection with the mutual information $I(\tilde{\mathbf{x}}; \mathbf{z})$, which quantifies how much information $\mathbf{z}$ preserves about $\tilde{\mathbf{x}}$:
\begin{equation}
    I(\tilde{\mathbf{x}}; \mathbf{z}) = \sum_{\tilde{\mathbf{x}}, \mathbf{z}} \mathbb{P}(\tilde{\mathbf{x}}, \mathbf{z}) \log \frac{\mathbb{P}(\tilde{\mathbf{x}}, \mathbf{z})}{\mathbb{P}(\tilde{\mathbf{x}})\, \mathbb{P}(\mathbf{z})}.
\end{equation}
The mutual information between the input and the latent variable can be lower-bounded as:
\begin{equation}
    I(\tilde{\mathbf{x}}; \mathbf{z}) \geq H(\tilde{\mathbf{x}}) - \mathbb{E}_{\tilde{\mathbf{x}}}\!\left[ \mathcal{L}_{\text{rec}}(\tilde{\mathbf{x}}) \right].
\end{equation}

By minimizing $\mathcal{L}_{\text{rec}}$, we tighten this lower bound, effectively preventing the latent variable $\mathbf{z}$ from collapsing into an uninformative representation and guaranteeing that veracity-related features are retained during the encoding process.

While mutual information ensures semantic preservation, the modality-specific noise $\bm{\epsilon}$ may still cause distribution shift. To mitigate this noise mismatch, we introduce a symmetric KL (SKL) divergence between the modality-specific posteriors, denoted as $q_c = q_\phi(\mathbf{z} \mid \tilde{\mathbf{x}}_c)$ and $q_g = q_\phi(\mathbf{z} \mid \tilde{\mathbf{x}}_g)$:
\begin{equation} 
\mathcal{L}_{\text{skl}} = \frac{1}{2} \left[ \mathrm{KL}(q_c \,\|\, q_g) + \mathrm{KL}(q_g \,\|\, q_c) \right]. 
\label{skl} 
\end{equation}

This term serves as a denoising regularizer that aligns the latent distributions. Let $\mathbf{W}$ denote the linear mapping matrix to the latent space. By expanding the SKL divergence for Gaussian posteriors, we derive a second-order lower bound (see Section~\ref{sec:theory_prove} for details):
\begin{equation}
\mathcal{L}_{\text{skl}} \geq \frac{1}{2} \left\| \mathbf{W}\bigl(\bm{\epsilon}_c - \bm{\epsilon}_g\bigr) \right\|^2_{\bm{\Sigma}_c^{-1}}.
\end{equation}
Crucially, since the shared semantic component $\mathbf{W}\mathbf{s}$ cancels out in the mean difference ($\bm{\mu}_c - \bm{\mu}_g$), this bound exclusively penalizes discrepancies induced by modality-specific noise. Minimizing $\mathcal{L}_{\text{skl}}$ thus effectively suppresses the noise inherent in virtual propagation by anchoring it to the relatively stable content semantics.

Following the denoising and alignment phase, we obtain the calibrated latent representations $\bm{\mu}_c$ and $\bm{\mu}_g$. To adaptively integrate these modalities, we employ a gated fusion mechanism that dynamically weighs their contributions based on the latent semantics:
\begin{align}
    a &= \sigma\left(\mathbf{W}_f [\bm{\mu}_c \oplus \bm{\mu}_g] + \mathbf{b}_f\right), \\
    \mathbf{o}_{f} &= a \cdot \bm{\mu}_c + (1 - a) \cdot \bm{\mu}_g,
\end{align}
where $\oplus$ denotes concatenation and $\sigma$ is the sigmoid activation function. The gate $a \in (0, 1)$ functions as a modality-wise attention that prioritizes the reliable content anchor when the simulated propagation exhibits high uncertainty or artifacts. The fused representation $\mathbf{o}_{f}$ is then fed into a classification network:
\begin{equation}
    \hat{y} = \textrm{Softmax}\left(\textrm{MLP}(\mathbf{o}_{f})\right).
\end{equation}

The final objective function $\mathcal{L}_{\text{total}}$ combines the cross-entropy classification loss $\mathcal{L}_{\text{cls}}$ with two VAE-driven regularizers that ensure semantic reconstruction and structural alignment:
\begin{equation}
\label{total_loss}
\mathcal{L}_{\text{total}} = \mathcal{L}_{\text{cls}} + \lambda_{\text{rec}} \cdot \mathcal{L}_{\text{rec}} + \lambda_{\text{skl}} \cdot \mathcal{L}_{\text{skl}},
\end{equation}
where $\lambda_{\text{rec}}$ and $\lambda_{\text{skl}}$ are hyperparameters that balance detection accuracy, information preservation, and noise suppression across modalities.

\subsubsection{Theoretical Justification of Denoising-Guided Alignment}
\label{sec:theory_prove}

This section provides theoretical justification for the two regularizers used in Section~\ref{denoise}. We show that (i) the reconstruction objective $\mathcal{L}_{\text{rec}}$ tightens a lower bound on the mutual information between the input $\mathbf{x}$ and the latent variable $\mathbf{z}$, preventing representation collapse, and (ii) the symmetric KL divergence (SKL) primarily suppresses modality-specific noise discrepancy rather than shrinking shared semantics.

\textit{Reconstruction as a mutual-information lower bound.}
We introduce a reconstruction loss $\mathcal{L}_{\text{rec}}$ to compel the latent variable $\mathbf{z}$ to retain essential semantics from the input $\mathbf{x}$. Minimizing this loss serves a dual role: it trains the decoder to reconstruct $\mathbf{x}$ from $\mathbf{z}$ while theoretically tightening a provable lower bound on their mutual information, thus preventing $\mathbf{z}$ from collapsing into an uninformative representation. Specifically,
\begin{align}
I(\mathbf{x}; \mathbf{z})
&= \mathbb{E}_{\mathbf{x}, \mathbf{z}} \left[ \log \frac{p(\mathbf{x} \mid \mathbf{z})}{p(\mathbf{x})} \right] \nonumber \\
&= \mathbb{E}_{\mathbf{x}, \mathbf{z}} \left[ \log \frac{p_\psi(\mathbf{x} \mid \mathbf{z})}{p(\mathbf{x})} \right]
+ \mathbb{E}_{\mathbf{x}, \mathbf{z}} \left[ \log \frac{p(\mathbf{x} \mid \mathbf{z})}{p_\psi(\mathbf{x} \mid \mathbf{z})} \right] \nonumber \\
&= \mathbb{E}_{\mathbf{x}, \mathbf{z}} \left[ \log \frac{p_\psi(\mathbf{x} \mid \mathbf{z})}{p(\mathbf{x})} \right]
+ \mathbb{E}_{\mathbf{z}} \Big[ \mathrm{KL}\!\left( p(\mathbf{x} \mid \mathbf{z}) \,\|\, p_\psi(\mathbf{x} \mid \mathbf{z}) \right) \Big] \nonumber \\
&\ge \mathbb{E}_{\mathbf{x}, \mathbf{z}} \left[ \log \frac{p_\psi(\mathbf{x} \mid \mathbf{z})}{p(\mathbf{x})} \right]
= -\mathbb{E}_{\mathbf{x}} \left[ \mathcal{L}_{\text{rec}}(\mathbf{x}) \right] + H(\mathbf{x}). \label{eq:mi_lb_full}
\end{align}

Therefore, minimizing $\mathcal{L}_{\text{rec}}$ tightens the lower bound in Eq.~\eqref{eq:mi_lb_full}, guaranteeing that $\mathbf{z}$ preserves essential information about $\mathbf{x}$.

\textit{Why $\mathcal{L}_{\text{rec}}$ prefers preserving shared semantics.} To investigate the noise suppression mechanism, we define the latent mean with a compression factor $\eta \in [0, 1]$ as $\bm{\mu}_{*}(\eta) = \eta\,\mathbf{W}\mathbf{s} + \mathbf{W}\mathbf{\epsilon}_{*}$. Under a Gaussian decoder assumption $\hat{\mathbf{x}}_{*} = \mathbf{W}^{-1}\mathbf{z}_{*}$, the reconstruction loss simplifies to an MSE objective. Decomposing the input as $\mathbf{x} \approx \mathbf{W}(\mathbf{s} + \mathbf{\epsilon}_{*})$, the loss with respect to $\eta$ becomes:
\begin{equation}
    \mathcal{L}_{\text{rec}}(\eta)
    = \frac{1}{2\sigma_x^2} \left\|\mathbf{x} - \hat{\mathbf{x}}_{*}\right\|^2 + \kappa
    = \frac{\|\mathbf{W}\mathbf{s}\|^2}{2\sigma_x^2} (1 - \eta)^2 + \kappa,
\label{eq:rec_compact}
\end{equation}
Here, $\kappa$ represents noise-related terms independent of $\eta$. Differentiating Eq.~\eqref{eq:rec_compact} reveals that $\frac{\partial^2 \mathcal{L}_{\text{rec}}}{\partial \eta^2} > 0$, indicating strict convexity with a global minimum at $\eta = 1$. In the context of the full objective (Eq.~\ref{total_loss}), since $\mathcal{L}_{\text{skl}}$ is insensitive to the semantic component in the mean difference, the quadratic penalty in $\mathcal{L}_{\text{rec}}$ acts as the primary regularizer ensuring $\mathbf{W}\mathbf{s}$ is preserved. This dynamic effectively filters out modality-specific noise $\mathbf{\epsilon}_{*}$ while retaining shared semantics.

\textit{Denoising through symmetric KL divergence.}
Given that semantic information is preserved via $\mathcal{L}_{\text{rec}}$, we next show how SKL suppresses residual noise discrepancy between the two latent posteriors.
For Gaussian posteriors $q_c=\mathcal{N}(\bm{\mu}_c,\bm{\Sigma}_c)$ and $q_g=\mathcal{N}(\bm{\mu}_g,\bm{\Sigma}_g)$, we derive the following lower bound:
\begin{align}
\operatorname{SKL}(q_c,q_g)
&=\frac12\Bigl[
      \operatorname{tr}\!\bigl(\bm{\Sigma}_c^{-1}\bm{\Sigma}_g+\bm{\Sigma}_g^{-1}\bm{\Sigma}_c\bigr)-2d
      +\Delta\bm{\mu}^{\top}\!\bigl(\bm{\Sigma}_c^{-1}+\bm{\Sigma}_g^{-1}\bigr)\Delta\bm{\mu}
  \Bigr] \nonumber \\
&\ge \frac12\Bigl[
      \operatorname{tr}\!\bigl(\bm{\Sigma}_g^{-1}\bm{\Sigma}_c+\bm{\Sigma}_c^{-1}\bm{\Sigma}_g\bigr)-2d
      +\Delta\bm{\mu}^{\top}\bm{\Sigma}_c^{-1}\Delta\bm{\mu}
  \Bigr] \nonumber \\
&=\frac12\Bigl[
      \operatorname{tr}\!\bigl(\bm{\Lambda}+\bm{\Lambda}^{-1}-2\mathbf{I}\bigr)
      +\|\Delta\bm{\mu}\|_{\bm{\Sigma}_c^{-1}}^{2}
  \Bigr] \nonumber \\
&\ge \frac12\,
      \|\Delta\bm{\mu}\|_{\bm{\Sigma}_c^{-1}}^{2}
 =  \frac12\, \|\bm{\mu}_c-\bm{\mu}_g\|_{\bm{\Sigma}_c^{-1}}^{2} \nonumber \\
&= \frac12\,
      \bigl\|\mathbf{W}\!\bigl(\mathbf{\epsilon}_c-\mathbf{\epsilon}_g\bigr)\bigr\|_{\bm{\Sigma}_c^{-1}}^{2},
\label{eq:skl_denoising}
\end{align}
where $\Delta \bm{\mu} = \bm{\mu}_c - \bm{\mu}_g$ denotes the difference between Gaussian means, $\bm{\Lambda} = \bm{\Sigma}_c^{-1/2} \bm{\Sigma}_g \bm{\Sigma}_c^{-1/2}$ characterizes the relative covariance structure, and $d$ is the dimensionality of the latent variable. Eq.~\eqref{eq:skl_denoising} reveals that the shared semantic component $\mathbf{s}$ cancels out perfectly in the mean difference. Consequently, minimizing SKL acts as a regularizer that specifically minimizes the distance between modality-specific noises $\mathbf{\epsilon}_c$ and $\mathbf{\epsilon}_g$, effectively filtering out simulation artifacts without compromising the shared semantics preserved by $\mathcal{L}_{\text{rec}}$.

\section{Experiments}
\label{sec_experiments}



\subsection{Experimental Settings}

\subsubsection{Datasets}
We use five real-world datasets collected from social media. The datasets are described as follows:

\textit{Content-Based Datasets.} Our study utilizes three distinct datasets: PolitiFact and GossipCop, both English datasets from FakeNewsNet \cite{shu2020fakenewsnet}, and the Chinese Weibo dataset \cite{jin2017multimodal}. PolitiFact and GossipCop comprise 741 and 10,350 news articles, respectively, collected from fact-checking websites and labeled as real or fake, with some articles including images. The Weibo dataset, collected from the Weibo platform, contains 4,707 Chinese news instances, each uniquely characterized by the presence of an accompanying image, which is crucial for multimodal analysis.

\textit{Propagation Datasets.} To compare AVOID with propagation-based methods, we additionally use two datasets, PolitiFact-P and GossipCop-P \cite{dou2021user} (renamed PolitiFact-P and GossipCop-P for differentiation). Each entry includes the full propagation trace of a news item. Dataset statistics are shown in Table~\ref{dataset}.
    \begin{table}[tbp]
    \centering
    \caption{The statistics of the datasets.}
    \label{dataset}             
    \begin{tabular}{lccccc}
    \toprule
    Dataset    & GossipCop & PolitiFact & Weibo & GossipCop-P & PolitiFact-P\\
    \midrule
    Total News & 10,350    & 741       & 4,704  & 5,683  & 489 \\
    Fake News     & 2,387     & 391        & 2,772 & 1,969  & 199 \\
    Real News & 7,963     & 350        & 1,935 & 3,714 & 290 \\
    Images     & 10,350    & 298        & 4,707 & 5,683   & 224  \\
    \bottomrule
    \end{tabular}
\end{table}


\subsubsection{Implementation Details}

We implement our proposed AVOID framework using the PyTorch library and conduct all experiments on an NVIDIA GeForce RTX 4090 GPU. For the news and comment encoding, we set the feature dimension to 768 to maintain consistency with the output of the pre-trained BERT-base~~\cite{ding2020bert} model. Specifically, news content is truncated to a maximum of 50 sentences with 25 tokens each, while user comments are limited to 15 sentences with 10 tokens per sentence. Regarding the agent-based simulator, we invoke the DeepSeek-V3-Chat~\cite{liu2024deepseek} API with the temperature and top-k values fixed at 0.7 and 0.9, respectively. The resulting propagation graph is processed by a two-layer GAT~\cite{velivckovic2017graph} with a hidden dimension of 128. For the training process, we employ the Adam optimizer~\cite{kingma2014adam} with an initial learning rate of $1e{-4}$ and a minibatch size of 32. We determine the optimal hyperparameters by evaluating the performance on the validation set, which is split alongside the training and test sets in a 7:1:2 ratio. To prevent overfitting, we apply an early stopping~\cite{yao2007early} strategy with a patience of 6 epochs, allowing for a maximum of 100 training epochs across all datasets.

\subsubsection{Baselines}
To evaluate the effectiveness of AVOID, we compare it against a wide range of state-of-the-art fake news detection models. To ensure a fair and rigorous comparison, all baseline methods are benchmarked under the same experimental environment. Specifically, our comparison encompasses three distinct categories of methods: (1) Content-based models that utilize news text and associated images; (2) LLM-enhanced methods that leverage large language models to facilitate prediction; and (3) Propagation-based methods that explicitly exploit the structural diffusion patterns of news.

The Content-Based fake news detection model:
\begin{itemize}
  \item BERT \cite{devlin2019bert}: is a pre-trained language model, and we fine-tune its last two layers for detection.
  \item HAN \cite{han}: uses news text to identify fake news by building a hierarchical attention network that captures both word-level and sentence-level features.
  \item dEFEND \cite{defend}: proposes an explainable fake news detection model, which utilizes body text and users’ comments to find k explainable comments to improve fake news detection.
  \item CAFE \cite{cafe}: constructs a cross-modal alignment module to transform different modals’ features into a shared semantic space, then evaluates the ambiguity between different modalities.
  \item HMCAN \cite{hmcan}: devises a hierarchical multi-modal attention network to learn a multi-modal news representation. 
  \item BREAK \cite{break}: models news text and images as a fully connected sentence–image graph, combining sequence and graph encoders with denoising to capture broad-range semantics for fake news detection.
  \item CSFND \cite{peng2024not}: devises an unsupervised fake news detection framework to capture the relationships between news semantic feature space and fake news decision space. 
  \item ALGM \cite{algm}: proposes a framework based on the Markov random field and fuses cross-modal features by ambiguity. 
  \item MIMoE-FND \cite{liu2025modality}: introduces a hierarchical mixture-of-experts framework that explicitly models text–image modality interactions via unimodal prediction agreement and semantic alignment and routes posts to specialized fusion experts for robust multimodal detection.
\end{itemize}

In addition to the traditional content-based baselines, we further compare the following LLM-based fake news detection models:
\begin{itemize}
  \item ARG \cite{hu2024bad}: studies how large language models can assist fake news detection and introduces an adaptive rationale guidance network, where a small model leverages LLM-generated rationales to improve veracity prediction.
  \item L-Defense \cite{l-defense}: leverages the wisdom of crowds to extract evidence and generate justifications via prompting LLM.
  \item GenFEND \cite{nan2024let}: uses large language models to role-play diverse user profiles, generate synthetic comments for each news piece, and aggregate these generated feedback signals to enhance fake news detection.
  \item SheepDog \cite{sheepdog}: proposes a style-robust text-based fake news detector that uses LLM-generated style-diverse news rewritings and content-focused veracity cues to make stable predictions under different writing styles.
  \item Adstyle \cite{park2025adversarial}: employs LLM-generated adversarial style conversions to augment training data and learns a text-based fake news detector that is robust to style-transfer attacks.
\end{itemize}

The Propagation-Based fake news detection model:
\begin{itemize}
  \item UGGCN \cite{kipf2016semi}: applies graph convolutional networks to model node representations on the news propagation graph and performs semi-supervised fake news classification by aggregating neighborhood information.
  \item Bi-GCN \cite{bian2020rumor}: employs bi-directional graph convolutional networks on both the original and reversed propagation graphs to capture rumor spread and refutation patterns for rumor detection on social media.
  \item GACL \cite{sun2022rumor}: uses graph adversarial contrastive learning on propagation graphs to learn robust news representations by jointly applying adversarial perturbations and contrastive objectives for rumor detection.
  \item UPFD \cite{dou2021user}: learns user preferences through their past engaged posts, and combines content with graph modeling.
  \item MFAN \cite{zheng2022mfan}: integrates textual, visual, and social graph features in one unified framework for rumor detection.
  \item PSGT \cite{zhu2024propagation}: designs a propagation structure-aware graph Transformer that filters out noisy user interactions and models multi-scale diffusion structures for robust and interpretable fake news detection.
\end{itemize}

\subsection{Performance Comparison}
We first benchmark the performance of AVOID against baselines for content-only early fake news detection, including text-based models, multimodal content-based methods, and LLM-enhanced approaches. Experiments are conducted on two English datasets (PolitiFact and GossipCop) and one Chinese dataset (Weibo) to comprehensively evaluate detection performance. The results are summarized in Table~\ref{tab:main_results}, and several key findings can be drawn as follows:

\begin{table*}[tbp]
  \centering
  \caption{Performance comparison between AVOID and baseline methods.
  Best results are in \textbf{bold}, and second best are \underline{underlined}.}
  \label{tab:main_results}

  \footnotesize
  \setlength{\tabcolsep}{3pt}
  \renewcommand{\arraystretch}{1.05}
  \newcolumntype{Y}{>{\centering\arraybackslash}X}

  \begin{tabularx}{\textwidth}{c|c|YYYY|YYYY|YYYY}
    \toprule
    \multirow{2}{*}{Category} & \multirow{2}{*}{Method}
      & \multicolumn{4}{c|}{PolitiFact}
      & \multicolumn{4}{c|}{GossipCop}
      & \multicolumn{4}{c}{Weibo} \\
    \cmidrule(lr){3-6} \cmidrule(lr){7-10} \cmidrule(lr){11-14}
    & & Acc. & Pre. & Rec. & F1
      & Acc. & Pre. & Rec. & F1
      & Acc. & Pre. & Rec. & F1 \\
    \midrule
    \multirow{9}{*}{Content-Based}
      & BERT      & 0.801 & 0.799 & 0.801 & 0.798
                  & 0.828 & 0.820 & 0.828 & 0.801
                  & 0.863 & 0.868 & 0.854 & 0.859 \\
      & HAN       & 0.857 & 0.859 & 0.854 & 0.856
                  & 0.837 & 0.823 & 0.837 & 0.831
                  & 0.863 & 0.863 & 0.855 & 0.858 \\
      & dEFEND    & 0.825 & 0.830 & 0.821 & 0.823
                  & 0.849 & 0.846 & 0.849 & 0.847
                  & 0.857 & 0.865 & 0.843 & 0.850 \\
      & CAFE      & 0.773 & 0.754 & 0.780 & 0.759
                  & 0.814 & 0.824 & 0.814 & 0.820
                  & 0.836 & 0.857 & 0.837 & 0.847 \\
      & HMCAN     & 0.894 & 0.898 & 0.892 & 0.896
                  & 0.836 & 0.818 & 0.836 & 0.820
                  & 0.896 & 0.912 & 0.881 & 0.890 \\
      & BREAK     & 0.920 & 0.921 & 0.920 & 0.921
                  & \underline{0.864} & \underline{0.861} & \underline{0.864} & \underline{0.863}
                  & 0.899 & 0.901 & 0.897 & 0.899 \\
      & CSFND     & 0.907 & 0.907 & 0.907 & 0.907
                  & 0.837 & 0.851 & 0.837 & 0.848
                  & 0.886 & 0.874 & 0.882 & 0.874 \\
      & ALGM      & 0.887 & 0.895 & 0.881 & 0.886
                  & 0.831 & 0.817 & 0.831 & 0.805
                  & 0.854 & 0.858 & 0.861 & 0.866 \\
      & MIMOE-FND & 0.878 & 0.892 & 0.883 & 0.886
                  & 0.854 & 0.847 & 0.854 & 0.847
                  & 0.912 & \underline{0.917} & \underline{0.918} & 0.909 \\
    \midrule
    \multirow{5}{*}{LLM-Enhanced}
      & ARG        & \underline{0.926} & \underline{0.924} & \underline{0.928} & \underline{0.924}
                   & 0.852 & 0.812 & 0.852 & 0.823
                   & 0.908 & 0.913 & 0.902 & 0.905 \\
      & L-Defense  & 0.893 & 0.902 & 0.899 & 0.896
                   & 0.862 & 0.858 & 0.862 & 0.861
                   & 0.875 & 0.874 & 0.875 & 0.875 \\
      & GenFEND    & 0.899 & 0.897 & 0.900 & 0.898
                   & 0.850 & 0.853 & 0.850 & 0.848
                   & 0.910 & 0.906 & 0.910 & 0.907 \\
      & SheepDog   & 0.913 & 0.912 & 0.909 & 0.911
                   & 0.843 & 0.831 & 0.843 & 0.834
                   & 0.905 & 0.908 & 0.902 & 0.909 \\
      & Adstyle    & 0.899 & 0.904 & 0.899 & 0.901
                   & 0.857 & 0.851 & 0.857 & 0.849
                   & \underline{0.918} & 0.913 & 0.917 & \underline{0.916} \\
    \midrule
    \multirow{2}{*}{Our Method}
      & \textbf{AVOID} & \textbf{0.960} & \textbf{0.956} & \textbf{0.960} & \textbf{0.958}
                       & \textbf{0.882} & \textbf{0.876} & \textbf{0.882} & \textbf{0.879}
                       & \textbf{0.936} & \textbf{0.930} & \textbf{0.932} & \textbf{0.934} \\
      & Imp.(\%)       & +3.67 & +3.46 & +3.44 & +3.68
                       & +2.08 & +1.74 & +2.08 & +1.85
                       & +1.96 & +1.42 & +1.53 & +1.86 \\
    \bottomrule
  \end{tabularx}
\end{table*}

\begin{itemize}
    \item AVOID consistently achieves the best overall performance across all datasets. As shown in Table~\ref{tab:main_results}, AVOID attains the highest scores on PolitiFact, GossipCop, and Weibo. Compared with the strongest baseline on each dataset, AVOID improves accuracy by 3.67\% on PolitiFact, 2.08\% on GossipCop, and 1.96\% on Weibo, with consistent gains in precision, recall, and F1. This advantage across three heterogeneous platforms indicates that AVOID generalizes effectively to diverse news domains and languages.

    \item AVOID outperforms traditional multimodal content-based baselines (e.g., CAFE, HMCAN). Unlike most baselines that directly fuse text and visual features for classification, AVOID leverages multimodal signals to drive agents’ decisions during virtual propagation. The results suggest that implicitly encoding multimodal evidence through the interaction patterns of the simulated propagation graph offers a more effective mechanism for detection than static feature fusion.

    \item While LLM-enhanced approaches generally surpass conventional detectors by using LLMs to generate auxiliary signals or perform direct reasoning, AVOID achieves further improvements. This indicates that merely using LLMs for semantic understanding is less effective than AVOID's strategy: exploiting LLM-empowered agents to simulate real-world social interactions. The superior performance validates that the interaction evidence derived from virtual propagation paths serves as crucial additional supervision.

    \item AVOID demonstrates the value of propagation-level social evidence over static comment augmentation. Specifically, while dEFEND utilizes real user comments and GenFEND employs LLM-generated comments, AVOID models dynamic social signals by simulating interactions among heterogeneous agents. The consistent performance superiority of AVOID over these baselines suggests that the structural and interaction patterns within the virtual propagation graph provide more informative and complementary cues for early fake news detection than individual comment features.
\end{itemize}

\subsection{Early-Stage Detection Evaluation}
To evaluate the performance of AVOID in the early stage of news diffusion, especially when only a very small portion of the real propagation is observable or even no propagation is available, we design an early detection experiment on PolitiFact-P and GossipCop-P.

Concretely, for each news piece, we sort all user interactions in chronological order and define three early-diffusion stages: (1) Content only (stage 0), we use only the news content itself and discard all real propagation interactions, simulating an extreme early detection setting with \emph{zero} observable propagation information; (2) 10\% stage, we truncate the real propagation to the earliest 10\% of interactions in time, simulating the stage where diffusion has just started, and the system can observe only a short segment of the cascade; (3) 30\% stage, we similarly truncate the real propagation to the earliest 30\% of interactions in time, simulating a stage where the news has diffused to some extent but is still in a relatively early phase.

Using the reconstructed propagation graphs, we conduct a comparative analysis between AVOID and representative baselines across different categories. Table~\ref{tab:prop_methods} summarizes their performance, from which we can observe the following:

\begin{table*}[tbp]
  \centering
  \caption{Performance of propagation-based methods on PolitiFact-P and Twitter with time-truncated propagation for early fake news detection. The best result in each column is highlighted in \textbf{bold} and the second best is \underline{underlined}.}
  \label{tab:prop_methods}
  \small
  \setlength{\tabcolsep}{3pt}
  \renewcommand{\arraystretch}{1.0}
  \newcolumntype{Y}{>{\centering\arraybackslash}X}

  \begin{tabularx}{\textwidth}{c c YYYYYYYYYYYYYY}
    \toprule
    \multirow{5}{*}{Category}
      & \multirow{5}{*}{Method}
      & \multicolumn{6}{c}{PolitiFact-P}
      & \multicolumn{6}{c}{GossipCop-P} \\
      \cmidrule(lr){3-8} \cmidrule(lr){9-14}
      & 
      & \multicolumn{2}{c}{0}
      & \multicolumn{2}{c}{0.1}
      & \multicolumn{2}{c}{0.3}
      & \multicolumn{2}{c}{0}
      & \multicolumn{2}{c}{0.1}
      & \multicolumn{2}{c}{0.3} \\
    \cmidrule(lr){3-4} \cmidrule(lr){5-6} \cmidrule(lr){7-8}
    \cmidrule(lr){9-10} \cmidrule(lr){11-12} \cmidrule(lr){13-14}
      &
      & Acc & F1 & Acc & F1 & Acc & F1
      & Acc & F1 & Acc & F1 & Acc & F1 \\
    \midrule

    \multirow{6}{*}{\shortstack{Propagation\\Enhanced}}
      & UDGCN  &   -   &   -   & 0.706 & 0.705 & 0.782 & 0.782
                 &   -   &   -   & 0.791 & 0.794 & 0.814 & 0.823 \\
      & BiGCN  &   -   &   -   & 0.661 & 0.632 & 0.777 & 0.774
                 &   -   &   -   & 0.824 & 0.822 & 0.841 & 0.836 \\
      & GACL   &   -   &   -   & 0.808 & 0.783 & 0.810 & 0.806
                 &   -   &   -   & 0.810 & 0.828 & 0.836 & 0.842 \\
      & UPFD   &   -   &   -   & 0.860 & 0.835 & \underline{0.886} & \underline{0.874}
                 &   -   &   -   & 0.824 & 0.858 & 0.867 & 0.881 \\
      & MFAN  &   -   &   -   & 0.842 & 0.825 & 0.872 & 0.860
                 &   -   &   -   & 0.824 & 0.865 & 0.864 & 0.881 \\
      & PSGT   &   -   &   -   & 0.863 & 0.852 & 0.872 & 0.868
                 &   -   &   -   & 0.869 & \underline{0.877} & \underline{0.881} & \underline{0.885} \\

    \midrule
    \multirow{3}{*}{\shortstack{Content-Based}}
    & BERT  &  0.784 & 0.788 & 0.784 & 0.788 & 0.784 & 0.788
                 & 0.804 & 0.816 & 0.804 & 0.816 & 0.804 & 0.816 \\
    & L-Defense   & \underline{0.880} & \underline{0.871} & \underline{0.880} & \underline{0.871} & 0.880 & 0.871
                 & 0.866 &  0.841  & 0.866 & 0.841 & 0.866 & 0.841 \\
    & BREAK   &  0.872 & 0.868 & 0.874 & 0.868 & 0.874 & 0.868
                 & \underline{0.878} & \underline{0.874}  & \underline{0.878} & 0.874 & 0.878 & 0.874 \\
    \midrule
    \multirow{2}{*}{Proposed}
      & \textbf{AVOID}
                 & \textbf{0.908} & \textbf{0.902}
                 & \textbf{0.908} & \textbf{0.902}
                 & \textbf{0.908} & \textbf{0.902}
                 & \textbf{0.893} & \textbf{0.896}
                 & \textbf{0.893} & \textbf{0.896}
                 & \textbf{0.893} & \textbf{0.896} \\
      & Imp.(\%) & +3.18 & +3.56 & +3.18 & +3.56 & +2.48 & +3.20
                 & +1.68 & +2.52 & +1.71 & +2.17 & +1.36 & +1.24 \\
    \bottomrule
  \end{tabularx}
\end{table*}

\begin{itemize}
    \item Under the early-diffusion setting, AVOID consistently achieves superior detection performance. On both PolitiFact-P and Twitter, whether we use only content (stage~0) or truncate the real propagation to the earliest 10\% or 30\% of interactions, AVOID attains the best Accuracy and F1 in every column and shows clear advantages over all propagation-based baselines. This indicates that, under the constraint of ``only observing early diffusion fragments'', AVOID is more suitable for early-stage fake news detection than methods that rely on complete real propagation paths.
    
    \item AVOID effectively mitigates the cold-start limitation by providing a usable substitute when real-world propagation is unavailable. In the extreme case where no observable cascade has emerged (Stage 0), propagation-enhanced baselines become inapplicable. However, AVOID consistently outperforms competitive content-based models by leveraging agent-simulated diffusion as a proxy for social feedback. This performance gain, achieved without access to real propagation samples, confirms that our framework remains highly valuable in the earliest stages of news dissemination, where it successfully distills veracity-related semantics from simulated interactions.
\end{itemize}

\subsection{Ablation Study}

To assess the distinct contribution of AVOID's components, we evaluate five variants against the full model. First, \textit{w/o filter} removes the confidence-based filtering stage and simulates a propagation path for every news piece rather than only for hard samples identified by the base detector. Second, \textit{w/o persona} operates without incorporating extracted persona information into the user profile. Third, \textit{w/o img} considers only the textual content of news articles, disregarding visual inputs. Fourth, \textit{w/o verifier} removes the specialized verified agents, retaining only diffuser agents for interaction. Finally, to test the feature alignment strategy, \textit{w/o skl} removes the SKL denoising module and instead directly concatenates textual and graph features.

\begin{figure*}[htbp]
    \centering
    \includegraphics[width=0.96\linewidth]{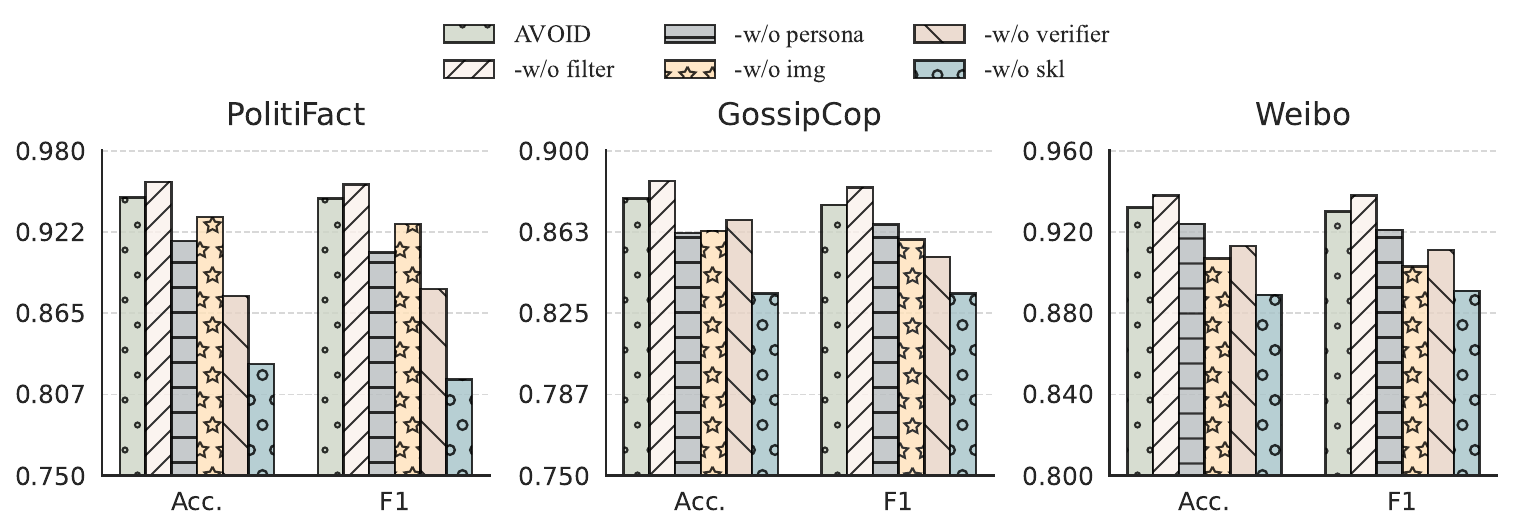}
    \caption{Ablation study of AVOID on three datasets.}
    \label{Ablation}
    \Description{Ablation}
\end{figure*}

As shown in Figure \ref{Ablation}, the ablation results validate the unique contribution of each component in AVOID. The detailed analysis is as follows:

\begin{itemize}
    \item Impact of filtering (\textit{w/o filter}): Removing the confidence-based filter and simulating propagation for every news piece slightly improves performance—by about 1.2\% on PolitiFact, 0.6\% on GossipCop, and 0.8\% on Weibo—but requires more than three times as many tokens. This indicates that our default AVOID configuration deliberately trades a small amount of accuracy for substantial efficiency gains; when computational resources are abundant, one may disable filtering and simulate propagation for all news to obtain the best possible performance.
    \item Necessity of Feature Alignment (\textit{w/o skl}): The removal of the SKL module caused a substantial performance drop, validating its critical role in mitigating modality discrepancies. Without SKL, direct concatenation of unaligned features introduces semantic noise, hindering robust decision-making.
    \item Role of Verified Agents (\textit{w/o verifier}): The absence of the verified agent mechanism led to a clear accuracy decline. Modeling high-credibility agents is crucial for guided propagation and effectively mitigating rumor spread.
    \item Benefit of User Personas (\textit{w/o persona}): Removing persona embeddings consistently lowered performance, demonstrating that using real user data provides valuable social context. Incorporating personalized characteristics proves an effective approach for comprehensively enhancing model robustness.
    \item Contribution of Image Features (\textit{w/o img}): Excluding image features resulted in a slight but noticeable performance drop. Visual information plays acomplementary role, offering additional signals when the model relies heavily on structural and textual cues.
\end{itemize}

\subsection{Comparison with Real-World Diffusion Cascades}

Although preliminary experiments verified AVOID's effectiveness in detection performance, this serves as indirect evidence regarding the fidelity of its generated propagation processes. To address this, we designed additional experiments to directly compare AVOID's simulated propagation paths with real cascades observed on social platforms, assessing the extent to which LLM-driven multi-agent simulations can reproduce real diffusion patterns.

Concretely, we compare real and simulated cascades from both structural and behavior-alignment perspectives. On the structural side, we characterize diffusion graphs using cascade depth, average node degree, edge density, structural virality (SV), and clustering coefficient, and additionally report the Jensen--Shannon Divergence (JSD) between degree distributions to capture distribution-level discrepancies beyond mean degree. On the behavioral side, we evaluate whether the simulation reproduces heterogeneous user reactions by reporting the verifier counts and the stance distribution (Pos/Neu/Neg) of comments to quantify opinion polarity at the population level. The quantitative results are summarized in Table~\ref{tab:graph_props} and Table~\ref{tab:behavior_props}, from which we draw the following conclusions:

\begin{table}[tbp]
\centering
\caption{Comparison of graph properties between virtual propagation paths and real data across two datasets.}
\label{tab:graph_props}
\begin{tabular}{@{}llcccccc@{}}
\toprule
\textbf{Dataset} & \textbf{Type} &
\textbf{Cascade Depth} & \textbf{Degree} &
\textbf{Density} & \textbf{SV} & \textbf{Cluster} & \textbf{JSD} \\
\midrule
\multirow{2}{*}{PolitiFact-P}
 & Real    & 4.203 & 1.884 & 0.016 & 2.808 & 0.114 & \multirow{2}{*}{0.078} \\
 & Virtual & 4.452 & 2.163 & 0.022 & 2.932 & 0.124 &                     \\
\midrule
\multirow{2}{*}{GossipCop-P}
 & Real    & 3.086 & 2.123 & 0.051 & 2.302 & 0.070 & \multirow{2}{*}{0.104} \\
 & Virtual & 3.221 & 2.635 & 0.043 & 2.186 & 0.065 &                     \\
\bottomrule
\end{tabular}
\end{table}

\begin{table}[tbp]
\centering
\caption{Comparison of behavioral properties between real and virtual cascades, including verifier frequency and comment stance distribution. Real verifiers are counted from verified-user metadata, while virtual verifiers are predefined agents; stances for both real and simulated comments are labeled with the same classifier.}
\label{tab:behavior_props}
\begin{tabular}{@{}lccc@{}}
\toprule
\textbf{Dataset} & \textbf{Type} & \textbf{Verifier} & \textbf{Pos:Neu:Neg (\%)} \\
\midrule
\multirow{2}{*}{PolitiFact-P}
 & real    & 4.88 & 54.6 : 9.3 : 36.1 \\
 & virtual & 5.00 & 57.8 : 10.7 : 31.5 \\
\midrule
\multirow{2}{*}{GossipCop-P}
 & real    & 2.29 & 75.8 : 18.4 : 5.8 \\
 & virtual & 2.00 & 73.2 : 22.1 : 3.7 \\
\bottomrule
\end{tabular}
\end{table}

\begin{itemize}
    \item AVOID reproduces key structural statistics of real cascades, including cascade depth, average degree, and edge density, suggesting that the simulated graphs match the empirical scale and connectivity patterns. The consistently low Jensen--Shannon divergence (JSD) further indicates that the overall distributions are well aligned. Together with the agreement in structural virality (SV), these results imply that AVOID captures the characteristic trade-off between breadth and depth that typically shapes real-world diffusion.
    \item Conditioning agents on fine-grained personas distilled from real comments improves behavioral alignment between simulated and observed cascades. This is supported by the close match in secondary-verifier counts and stance distributions between real and virtual comments, indicating that the simulation reflects heterogeneous reaction patterns rather than producing uniform, behavior-agnostic interactions.
    \item PolitiFact-P and GossipCop-P exhibit distinct diffusion dynamics and user behavior profiles in the real data, and AVOID maintains these dataset-specific differences in the simulated cascades. This suggests that the framework is sensitive to platform- and dataset-dependent discussion norms, avoids collapsing to a one-size-fits-all propagation template, and enables more faithful cross-domain simulation.
\end{itemize}


\subsection{Sensitivity of Hyperparameters}

\begin{figure}[htbp]
    \centering
    \includegraphics[width=1\linewidth]{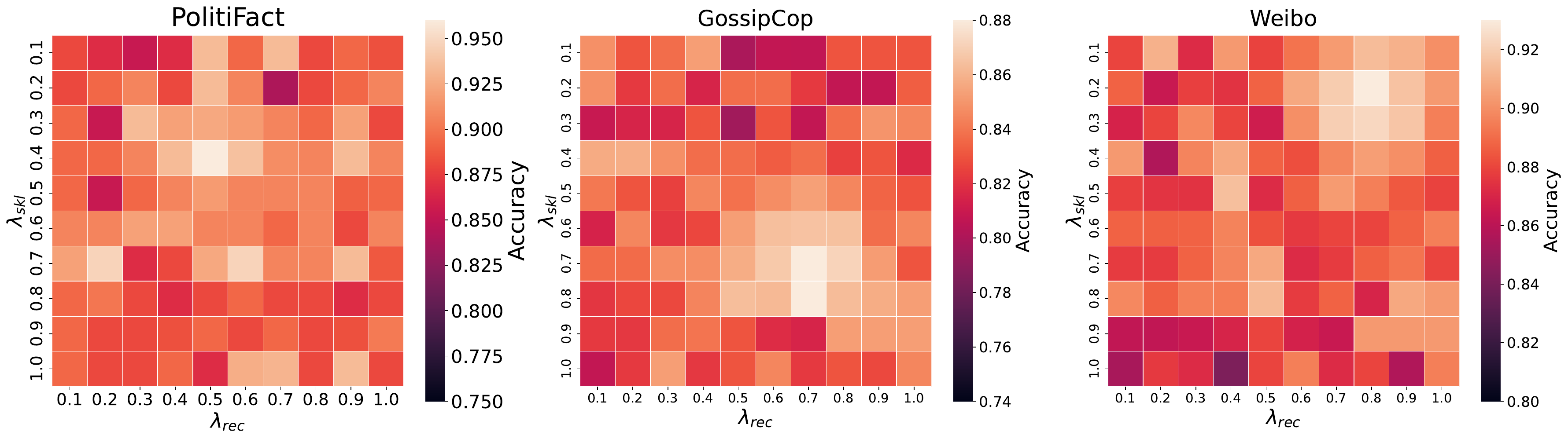}
    \caption{The effect of different hyparameters $\lambda_{\text{rec}}$ and $\lambda_{\text{skl}}$ on three datasets.}
    \label{fig6}
    \Description{hyperparam}
\end{figure}

In the AVOID model, the reconstruction loss $\mathcal{L}_{\text{rec}}$ is weighted by $\lambda_{\text{rec}}$ and controls the fidelity of unimodal feature reconstruction, while the SKL alignment loss $\mathcal{L}_{\text{skl}}$ is weighted by $\lambda_{\text{skl}}$ and determines the strength of cross-modal semantic consistency regularization. To investigate how these two components jointly affect detection performance, we conduct a grid search over $\lambda_{\text{rec}},\lambda_{\text{skl}} \in \{0.1, 0.2, \ldots, 1.0\}$ with a step size of 0.1, keeping all other hyperparameters fixed. For each $(\lambda_{\text{rec}},\lambda_{\text{skl}})$ combination, we train AVOID on the corresponding dataset and record the resulting detection accuracy, which we visualize as the heatmaps in Figure~\ref{fig6}.

Across the three datasets, the optimal values of the reconstruction loss $\mathcal{L}_{\text{rec}}$ and the SKL alignment loss $\mathcal{L}_{\text{skl}}$ are not identical, but they follow a consistent trend: they shift systematically with the strength of cross-modal coupling and the noise level of the data, indicating that this pair of hyperparameters exhibits good cross-dataset generalization. Concretely:

\begin{itemize}
    \item On PolitiFact, textual topics are focused, and images mostly play a supporting role. The model achieves its best performance around moderate weights (e.g., $\lambda_{\text{rec}}=0.5,\lambda_{\text{skl}}=0.4$), suggesting that modest cross-modal regularization is sufficient and further strengthening the alignment brings limited benefit. In contrast, on GossipCop the best performance appears at larger weights (around $\lambda_{\text{rec}}=0.7,\lambda_{\text{skl}}=0.7$), which reflects that in subjective, entertainment-oriented rumors with diverse propagation patterns, both stronger unimodal reconstruction and stronger cross-modal consistency are needed to suppress semantic uncertainty between text and images.
    \item On Weibo, the model attains its best accuracy at $(\lambda_{\text{rec}}=0.8,\lambda_{\text{skl}}=0.2)$, and remains competitive in a surrounding region with moderately large $\lambda_{\text{rec}}$ and moderate $\lambda_{\text{skl}}$ (approximately $\lambda_{\text{rec}}\in[0.5,0.8], \lambda_{\text{skl}}\in[0.2,0.6]$). This suggests that strong unimodal reconstruction is particularly beneficial on this dataset, while only moderate cross-modal alignment is needed. The pattern is consistent with short, noisy microblog posts where images serve highly diverse purposes (e.g., emojis, decorative or meme-style pictures): if the alignment constraint is pushed too aggressively, such loosely related visual signals may be forced into the shared semantic space and introduce additional noise.
    \item Since $\lambda_{\text{rec}}$ controls unimodal robustness and $\lambda_{\text{skl}}$ governs inter-modal semantic calibration/denoising, AVOID's $(\lambda_{\text{rec}}, \lambda_{\text{skl}})$ parameters should not be globally fixed. They must be adaptively tuned based on the specific dataset's inherent text-image coupling strength and noise characteristics.
\end{itemize}

\subsection{Token Usage}

To better understand the resource efficiency of AVOID, we further compare its token usage with other LLM-based fake news detection methods. Specifically, for each model, we record the total number of tokens consumed when evaluating the entire test set (including both prompts and generated outputs) as a proxy for inference cost, and report this token budget alongside the corresponding detection accuracy to examine the trade-off between performance and efficiency.

\begin{figure*}[tbp]
    \centering
    \includegraphics[width=0.92\linewidth]{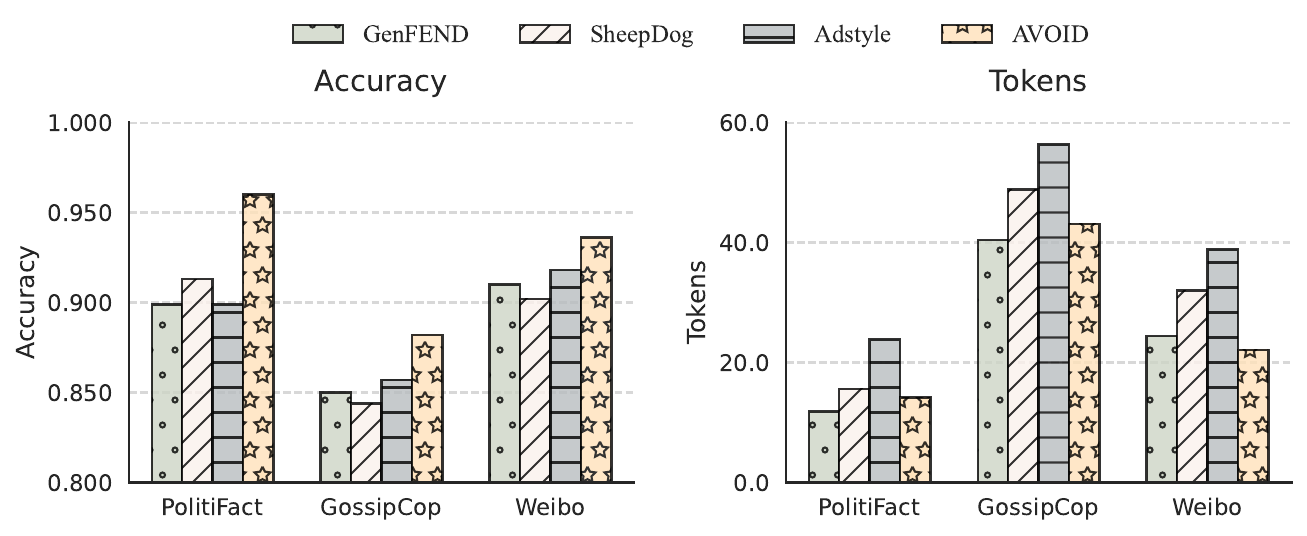}
    \caption{Comparative results of various LLM-based fake news detection model of accuracy and token consumption across three datasets.}
    \label{token usage}
    \Description{token usage}
\end{figure*}

\begin{itemize}
    \item Across all three datasets, AVOID consistently achieves the highest detection accuracy among LLM-based baselines, while its token consumption remains relatively low compared to other methods. In the Weibo dataset,  AVOID is also the most token-efficient model, showing that its performance gains do not rely on simply spending more tokens.
    \item Since AVOID only triggers multi-agent propagation simulation on items that pass a difficulty-based filter, and handles easy cases with cheaper content-only inference, it avoids the quadratic cost of simulating full cascades for every instance. This selective strategy improves early detection performance while keeping the overall token usage moderate, offering a practically useful balance between effectiveness and resource consumption.
\end{itemize}



\subsection{Case Study}

We evaluate AVOID through two complementary case studies. First, a micro-level analysis of agent reasoning demonstrates the generation mechanism. Second, a macro-level comparison between simulated and real cascades confirms the structural fidelity of the generated diffusion.

\subsubsection{Micro-level Analysis: Agent Reasoning and Generation}

\begin{figure*}[htbp]
    \centering
    \includegraphics[width=0.98\linewidth]{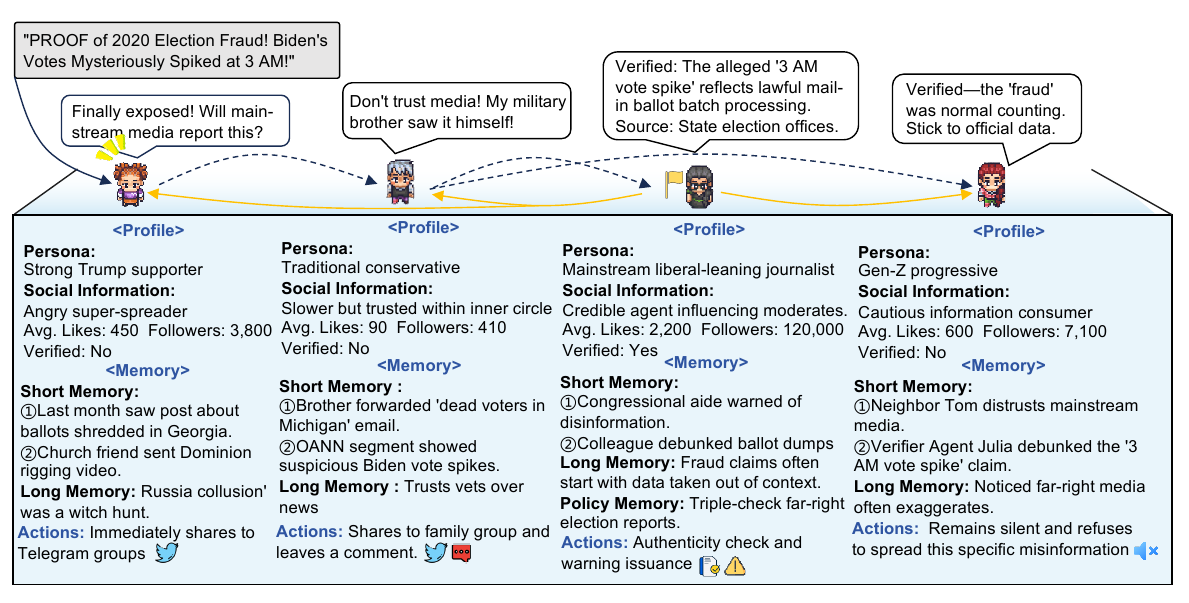}
    \caption{Case study of simulated misinformation propagation, showing how agents with different personas, memories, and social traits respond and influence the spread.}
    \label{casestudy}
    \Description{case study.}
\end{figure*}

Figure~\ref{casestudy} visualizes the lifecycle of a simulated propagation path generated by AVOID, demonstrating how the cognition of the individual agent translates into the dynamics of collective diffusion. The simulation initiates by instantiating agents with fine-grained personas distilled from real-world data. As shown in the figure, an agent is initialized with the profile of a "Strong Trump supporter," inheriting specific ideological priors and behavioral patterns. This data-driven grounding ensures that the simulation starts with agents who possess realistic social attributes rather than generic, stochastic behaviors.

Upon encountering the target misinformation (e.g., the "3 AM vote spike" claim), the agent triggers a retrieval-augmented reasoning process. By consulting its long-term memory, the agent retrieves contextually related beliefs—such as the view that "Russia collusion was a witch hunt." This retrieval reinforces the agent's confirmation bias, aligning the new claim with its existing worldview. Consequently, this cognitive coherence catalyzes the decision to "Immediately share" the content, reflecting a plausible psychological reaction to the stimulus.

Finally, the aggregation of these individual micro-decisions creates the macro-level structure of information flow. Each interaction forms a directed edge, culminating in a complete virtual propagation graph. This generated graph serves as a structural proxy for the missing real-world data, providing the downstream detector with the rich, dynamic social context necessary for robust early detection.

\subsubsection{Macro-level Analysis: Diffusion Fidelity and Graph Structure}

\begin{figure*}[htbp]
    \centering
    \includegraphics[width=0.75\linewidth, height=0.52\linewidth]{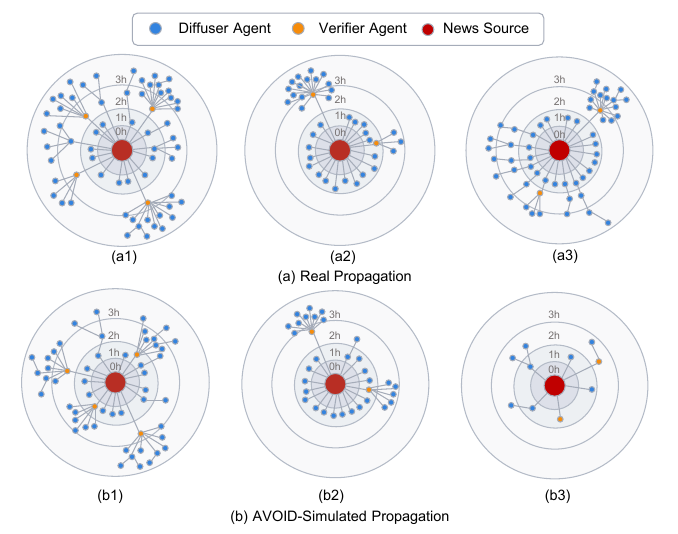}
    \caption{The first row (a1--a3) shows the real propagation cascades, while the second row (b1--b3) shows the AVOID Simulated cascades.}
    \label{casestudy_comparison}
    \Description{case study.}
\end{figure*}

To strengthen our macro-level evidence on diffusion fidelity, we visualize AVOID-simulated cascades alongside their real-world counterparts. As illustrated in Figure~\ref{casestudy_comparison}, we select representative fake news instances from the PolitiFact and GossipCop datasets to show that our framework preserves salient empirical structural signatures.

In the first two columns of Figure~\ref{casestudy_comparison}, AVOID shows strong agreement with the real propagation patterns. Across both PolitiFact and GossipCop, the simulated cascades (b1, b2) mirror the corresponding ground-truth graphs (a1, a2) in key structural aspects, including comparable branching depth and the emergence of local clusters around intermediate high-degree spreaders. This alignment supports the view that grounding agent behaviors in real-world personas and memory retrieval can reproduce the social interaction patterns that shape misinformation diffusion.

For the third example (PolitiFact~14448), the simulated cascade is much smaller than the real one. The claim is: \textit{inflammatory conspiracy narrative—alleging an Obama-led coup, “New World Order” symbolism, secret tunnel meetings, and a “population reduction” plot targeting Trump supporters.} Under AVOID’s persona- and memory-driven decision process, content that is both plainly implausible and overtly inciting is treated as not worth amplifying, so the diffusion quickly dies out and only a handful of agents participate (b3). This behavior is also consistent with the safety alignment of the underlying large language model, which further discourages agents from spreading highly sensational, harm-framed narratives.

\section{Related Work}
\label{sec_related}

\subsection{Fake News Detection}
\label{fakenews}
\subsubsection{Content-centric early fake news detection.}
Early approaches to fake news detection primarily focused on the textual content of news articles. These methods employed word embeddings in conjunction with convolutional or recurrent neural networks to learn representations of article text \cite{textcnn, han}. The advent of pre-trained language models marked a significant advancement. Subsequent research leveraged BERT-based \cite{devlin2019bert} encoders to generate contextualized representations, capturing richer semantic information within news content and achieving substantial performance improvements \cite{ding2020bert, kaliyar2021fakebert}. Building upon this, researchers explored techniques for modeling long-range dependencies within articles. For instance, BREAK \cite{break} jointly models sentence-level graph structures and sequential information to obtain more discriminative content representations, further mitigating both structural and feature-level noise. In parallel, some research has expanded the scope of content modeling by incorporating visual modalities alongside text. Representative multimodal methods \cite{cafe,hmcan,liu2025modality, peng2024not} jointly encode textual and visual features to learn cross-modal interactions. By fusing these heterogeneous signals, such models aim to capture complementary evidence and leverage cross-modal correlations for more robust veracity prediction.

\subsubsection{LLM-Enhanced early fake news detection.}

LLMs have introduced strong language understanding and reasoning capabilities, reshaping misinformation detection. However, direct zero-shot usage is often unreliable for fine-grained veracity judgment \cite{pelrine2023towards, zhou2023synthetic}. To mitigate this, early work \cite{pavlyshenko2023analysis} explored supervised adaptation and LLM-assisted detection, where LLMs provide multi-perspective rationales but may struggle to reliably select and aggregate evidence into a final decision; subsequent methods distill or integrate such rationales to guide smaller detectors in cost-sensitive settings \cite{l-defense, hu2024bad}. Alongside effectiveness, robustness has also become a key concern: LLM-driven stylistic reframing can camouflage fake news and undermine style-reliant detectors, motivating style-agnostic or attribution-focused designs that emphasize content veracity \cite{sheepdog, park2025adversarial}.

Beyond relying on parametric knowledge of LLMs, recent research increasingly shifts to evidence-based verification, where models explicitly interact with external sources during reasoning. Retrieval-Augmented Generation integrates retrieval with generation to obtain up-to-date evidence from search engines or knowledge bases, reducing hallucinations and mitigating knowledge cutoff issues \cite{lewis2020retrieval, factllamallama}. Building on this paradigm, decomposition-based methods further split complex claims into verifiable sub-questions and aggregate sub-results for finer-grained assessment \cite{hu2025decomposition}, while tool-augmented frameworks enable multi-step verification with auditable evidence trails; for instance, Self-Checker provides plug-and-play modules and a policy agent that plans verification actions and outputs verifiable rationales \cite{li2024self}.

Unlike these “single-model” enhancements, agent-based approaches coordinate multiple LLM agents for collaborative verification via role specialization, debate, and cross-checking. Debate-to-Detect \cite{han2025debate} reformulates misinformation detection as a structured multi-stage adversarial debate, assigning domain-specific profiles to agents and producing an interpretable verdict with a multi-dimensional judging rubric. MARO \cite{li2025multi} targets cross-domain misinformation detection by decomposing analysis into multiple expert agents, further introducing a question-reflection mechanism and automated decision-rule optimization to improve generalization. LoCal \cite{ma2025local} proposes an LLM-based multi-agent fact-checking framework that emphasizes logical and causal consistency, combining a decomposing agent, specialized reasoning agents, and evaluating agents to iteratively verify complex claims.

\subsubsection{Social context and propagation-based detection.}
Beyond content, many methods exploit social context, such as user comments and interactions as complementary signals for veracity prediction \cite{defend}. Another influential direction models diffusion trajectories as propagation graphs: users or posts are treated as nodes and repost, reply, comment relations form edges, enabling graph-based or temporal aggregation modules to capture rumor-related structural patterns for detection \cite{bian2020rumor, sun2022rumor, dou2021user, zhu2024propagation}. However, such propagation signals are often sparse or entirely unavailable at the early stage, making graph-based detectors unreliable in cold-start scenarios. To alleviate this limitation, recent studies treat propagation generation as a graph synthesis task: instead of merely extracting structural patterns from observed cascades, they learn generative models that can sample plausible propagation graphs or diffusion paths and use these synthetic structures as additional propagation evidence to strengthen downstream rumor detectors under cold-start settings \cite{hou2024dag, zhang2024mitigating, liu2020fned}. In addition, some methods \cite{nan2024let, wang2025collaboration} address early-stage comment scarcity by using LLM-based generators, often with persona diversification, to produce realistic discussion-style comments that can be incorporated as complementary social evidence for enhancing early rumor detection when real interactions are limited.

\subsection{LLM-Based Agents for Social Simulation}
\label{llm}
Integrating LLMs into agent-based social simulation has emerged as a promising direction for studying human behaviors and their collective dynamics. Early studies suggest that LLMs can reproduce outcomes of classical experiments in economics, psycholinguistics, and social psychology via carefully prompted role-play and interaction protocols \cite{aher2023using, horton2023large, xie2024can}. Building on this, recent work has moved from isolated decision tasks to large-scope simulation settings, where multiple LLM-driven agents inhabit interactive environments to generate long-horizon daily activities with memory and reflection, exemplified by generative agent societies and related simulators \cite{park2023generative,ju2025trajllm,wang2025yulan}. Beyond open-ended virtual societies, LLM agents have also been adopted as domain-grounded user surrogates for platform-scale ecosystems; for instance, RecAgent models realistic user behavior to support recommender-system simulation and intervention studies \cite{wang2025user}. Complementarily, controlled social game testbeds like repeated games and social dilemmas, provide a principled way to probe strategic interaction, cooperation, and coordination among LLM agents \cite{xu2023exploring, light2023avalonbench}. 

In particular, social media provides a natural testbed for such validation, motivating recent work to extend LLM-based social simulation to social network settings where LLM-empowered agents are situated on explicit interaction graphs to model networked interactions and information diffusion. For instance, Gao et al.~\cite{gao2023s3} proposed S$^3$, which instantiates LLM-driven users on real-world network data and explicitly simulates three key behavioral facets—emotion, attitude, and interaction—so that population-level phenomena (e.g., the spread of information, attitudes, and emotions) can emerge from individual-level perception and actions. In parallel, Liu et al.~\cite{liu2024skepticism} introduced an LLM-based fake-news propagation simulation framework in which each agent is conditioned on personality and equipped with short-/long-term memory and reflection, enabling day-by-day opinion exchange, attitude updating, and the evaluation of intervention strategies along the trajectory from skepticism to acceptance. Some works instead hybridize classical diffusion processes with LLM-based components to improve both controllability and scalability: they retain a standard diffusion mechanism as the backbone, while using LLM agents to generate language-level interactions and decision signals for a subset of key users and relying on lightweight rule-based/deductive agents for the remaining population. This design makes the propagation dynamics easier to interpret and tune, while keeping the simulation cost manageable \cite{mou2024unveiling, hu2024llm}. In addition, several works \cite{hu2025simulating, liu2024tiny} have investigated rumor or misinformation spreading with LLM-agent frameworks across different network structures and heterogeneous agent types, using such simulations to analyze how graph topology and persona-dependent behaviors jointly shape propagation outcomes.


\section{Conclusion}
\label{sec_conclusion}
This paper introduces AVOID, a novel framework for early fake news detection. To address the scarcity of real propagation data in early stages, AVOID constructs LLM-based agents whose personas are aligned to real user data and simulates plausible news propagation paths through multi-agent interactions. To better match real-world diffusion, we (i) differentiate verifier agents and diffuser agents to reflect heterogeneous engagement behaviors, and (ii) ground each agent’s actions in persona-conditioned decision rules, so that the induced interaction patterns are consistent with user-level traits observed in the data. These virtual paths provide crucial social context to supplement content features. Furthermore, we propose a propagation-aware fusion module with symmetric KL divergence to mitigate simulation noise by aligning the latent distributions of content and propagation representations. Experiments on real-world datasets show that AVOID significantly outperforms state-of-the-art baselines, demonstrating its effectiveness for early fake news detection.

Looking ahead, several promising directions remain to be explored to enhance the depth and utility of diffusion simulation. While the current framework captures propagation dynamics between typical users and opinion leaders, real-world information ecosystems are inherently adversarial. A critical avenue for future research is to incorporate adversarial agents that model the strategies of malicious actors, such as content fabrication and propagation manipulation, so as to better characterize the full lifecycle of disinformation from production to containment. In addition, evolving the simulator into a counterfactual testbed for policy intervention represents a significant opportunity. By enabling systematic counterfactual analysis of mitigation scenarios, the framework could provide principled guidance for proactive defense design, shifting the focus from passive observation to strategic ecosystem governance.

\begin{acks}

We thank the anonymous reviewers for their insightful feedback and constructive comments. This work was supported by the National Natural Science Foundation of China (Grant No. 62176028) and the Australian Research Council under the streams of Discovery Early Career Research Award (Grant No. DE250100613).
\end{acks}

\bibliographystyle{ACM-Reference-Format}
\bibliography{main}

@String{Computing = "Computing" }

@String{Computer = "{IEEE} Computer" }

@String{Springer = "Springer-Verlag" }

@article{allcott2017social,
  title={Social media and fake news in the 2016 election},
  author={Allcott, Hunt and Gentzkow, Matthew},
  journal={Journal of economic perspectives},
  volume={31},
  number={2},
  pages={211--236},
  year={2017},
  publisher={American Economic Association 2014 Broadway, Suite 305, Nashville, TN 37203-2418}
}

@article{roozenbeek2020susceptibility,
  title={Susceptibility to misinformation about COVID-19 around the world},
  author={Roozenbeek, Jon and Schneider, Claudia R and Dryhurst, Sarah and Kerr, John and Freeman, Alexandra LJ and Recchia, Gabriel and Van Der Bles, Anne Marthe and Van Der Linden, Sander},
  journal={Royal Society open science},
  volume={7},
  number={10},
  pages={201199},
  year={2020},
  publisher={The Royal Society}
}

@article{fisher,
  title={Pizzagate: From rumor, to hashtag, to gunfire in DC},
  author={Fisher, Marc and Cox, John Woodrow and Hermann, Peter},
  journal={Washington Post},
  volume={6},
  pages={8410--8415},
  year={2016}
}

@mastersthesis{textcnn,
  title={Convolutional neural network for sentence classification},
  author={Chen, Yahui},
  year={2015},
  school={University of Waterloo}
}

@inproceedings{zhang2021mining,
  title={Mining dual emotion for fake news detection},
  author={Zhang, Xueyao and Cao, Juan and Li, Xirong and Sheng, Qiang and Zhong, Lei and Shu, Kai},
  booktitle={Proceedings of the web conference 2021},
  pages={3465--3476},
  year={2021}
}

@inproceedings{cafe,
  title={Cross-modal ambiguity learning for multimodal fake news detection},
  author={Chen, Yixuan and Li, Dongsheng and Zhang, Peng and Sui, Jie and Lv, Qin and Tun, Lu and Shang, Li},
  booktitle={Proceedings of the ACM web conference 2022},
  pages={2897--2905},
  year={2022}
}

@inproceedings{defend,
  title={defend: Explainable fake news detection},
  author={Shu, Kai and Cui, Limeng and Wang, Suhang and Lee, Dongwon and Liu, Huan},
  booktitle={Proceedings of the 25th ACM SIGKDD international conference on knowledge discovery \& data mining},
  pages={395--405},
  year={2019}
}

@inproceedings{han,
  title={Hierarchical attention networks for document classification},
  author={Yang, Zichao and Yang, Diyi and Dyer, Chris and He, Xiaodong and Smola, Alex and Hovy, Eduard},
    editor = "Knight, Kevin  and
      Nenkova, Ani  and
      Rambow, Owen",
    booktitle = "Proceedings of the 2016 Conference of the North {A}merican Chapter of the Association for Computational Linguistics: Human Language Technologies",
    month = jun,
    year = "2016",
    address = "San Diego, California",
    publisher = "Association for Computational Linguistics",
    url = "https://aclanthology.org/N16-1174/",
    doi = "10.18653/v1/N16-1174",
    pages = "1480--1489"
}

@inproceedings{break,
  title={Graph with Sequence: Broad-Range Semantic Modeling for Fake News Detection},
  author={Yin, Junwei and Gao, Min and Shu, Kai and Li, Wentao and Huang, Yinqiu and Wang, Zongwei},
  booktitle={Proceedings of the ACM on Web Conference 2025},
  pages={2838--2849},
  year={2025}
}

@inproceedings{hmcan,
  title={Hierarchical multi-modal contextual attention network for fake news detection},
  author={Qian, Shengsheng and Wang, Jinguang and Hu, Jun and Fang, Quan and Xu, Changsheng},
  booktitle={Proceedings of the 44th international ACM SIGIR conference on research and development in information retrieval},
  pages={153--162},
  year={2021}
}

@inproceedings{algm,
  title={A Generalized Deep Markov Random Fields Framework for Fake News Detection.},
  author={Dong, Yiqi and He, Dongxiao and Wang, Xiaobao and Li, Yawen and Su, Xiaowen and Di Jin 0001},
  booktitle={IJCAI},
  pages={4758--4765},
  year={2023}
}

@article{llama,
  title={Llama: Open and efficient foundation language models},
  author={Touvron, Hugo and Lavril, Thibaut and Izacard, Gautier and Martinet, Xavier and Lachaux, Marie-Anne and Lacroix, Timoth{\'e}e and Rozi{\`e}re, Baptiste and Goyal, Naman and Hambro, Eric and Azhar, Faisal and others},
  journal={arXiv preprint arXiv:2302.13971},
  year={2023}
}

@inproceedings{l-defense,
  title={Explainable fake news detection with large language model via defense among competing wisdom},
  author={Wang, Bo and Ma, Jing and Lin, Hongzhan and Yang, Zhiwei and Yang, Ruichao and Tian, Yuan and Chang, Yi},
  booktitle={Proceedings of the ACM Web Conference 2024},
  pages={2452--2463},
  year={2024}
}

@inproceedings{sheepdog,
  title={Fake News in Sheep's Clothing: Robust Fake News Detection Against LLM-Empowered Style Attacks},
  author={Wu, Jiaying and Guo, Jiafeng and Hooi, Bryan},
  booktitle={Proceedings of the 30th ACM SIGKDD conference on knowledge discovery and data mining},
  pages={3367--3378},
  year={2024}
}

@article{vosoughi2018spread,
  title={The spread of true and false news online},
  author={Vosoughi, Soroush and Roy, Deb and Aral, Sinan},
  journal={science},
  volume={359},
  number={6380},
  pages={1146--1151},
  year={2018},
  publisher={American Association for the Advancement of Science}
}

@inproceedings{aher2023using,
  title={Using large language models to simulate multiple humans and replicate human subject studies},
  author={Aher, Gati V and Arriaga, Rosa I and Kalai, Adam Tauman},
  booktitle={International conference on machine learning},
  pages={337--371},
  year={2023},
  organization={PMLR}
}

@article{gao2023s3,
  title={S3: Social-network simulation system with large language model-empowered agents},
  author={Gao, Chen and Lan, Xiaochong and Lu, Zhihong and Mao, Jinzhu and Piao, Jinghua and Wang, Huandong and Jin, Depeng and Li, Yong},
  journal={arXiv preprint arXiv:2307.14984},
  year={2023}
}

@article{chen2024large,
  title={When large language models meet personalization: Perspectives of challenges and opportunities},
  author={Chen, Jin and Liu, Zheng and Huang, Xu and Wu, Chenwang and Liu, Qi and Jiang, Gangwei and Pu, Yuanhao and Lei, Yuxuan and Chen, Xiaolong and Wang, Xingmei and others},
  journal={World Wide Web},
  volume={27},
  number={4},
  pages={42},
  year={2024},
  publisher={Springer}
}

@article{liu2024skepticism,
  title={From skepticism to acceptance: Simulating the attitude dynamics toward fake news},
  author={Liu, Yuhan and Chen, Xiuying and Zhang, Xiaoqing and Gao, Xing and Zhang, Ji and Yan, Rui},
  journal={arXiv preprint arXiv:2403.09498},
  year={2024}
}

@article{paul2021deep,
  title={Deep learning on a data diet: Finding important examples early in training},
  author={Paul, Mansheej and Ganguli, Surya and Dziugaite, Gintare Karolina},
  journal={Advances in neural information processing systems},
  volume={34},
  pages={20596--20607},
  year={2021}
}

@article{sorscher2022beyond,
  title={Beyond neural scaling laws: beating power law scaling via data pruning},
  author={Sorscher, Ben and Geirhos, Robert and Shekhar, Shashank and Ganguli, Surya and Morcos, Ari},
  journal={Advances in Neural Information Processing Systems},
  volume={35},
  pages={19523--19536},
  year={2022}
}

@article{mou2024unveiling,
  title={Unveiling the truth and facilitating change: Towards agent-based large-scale social movement simulation},
  author={Mou, Xinyi and Wei, Zhongyu and Huang, Xuanjing},
  journal={arXiv preprint arXiv:2402.16333},
  year={2024}
}

@article{pelrine2023towards,
  title={Towards reliable misinformation mitigation: Generalization, uncertainty, and gpt-4},
  author={Pelrine, Kellin and Imouza, Anne and Thibault, Camille and Reksoprodjo, Meilina and Gupta, Caleb and Christoph, Joel and Godbout, Jean-Fran{\c{c}}ois and Rabbany, Reihaneh},
  journal={arXiv preprint arXiv:2305.14928},
  year={2023}
}

@article{achiam2023gpt,
  title={Gpt-4 technical report},
  author={Achiam, Josh and Adler, Steven and Agarwal, Sandhini and Ahmad, Lama and Akkaya, Ilge and Aleman, Florencia Leoni and Almeida, Diogo and Altenschmidt, Janko and Altman, Sam and Anadkat, Shyamal and others},
  journal={arXiv preprint arXiv:2303.08774},
  year={2023}
}

@inproceedings{hu2024bad,
  title={Bad actor, good advisor: Exploring the role of large language models in fake news detection},
  author={Hu, Beizhe and Sheng, Qiang and Cao, Juan and Shi, Yuhui and Li, Yang and Wang, Danding and Qi, Peng},
  booktitle={Proceedings of the AAAI Conference on Artificial Intelligence},
  volume={38},
  pages={22105--22113},
  year={2024}
}

@inproceedings{factllamallama,
  title={Factllama: Optimizing instruction-following language models with external knowledge for automated fact-checking},
  author={Cheung, Tsun-Hin and Lam, Kin-Man},
  booktitle={2023 Asia Pacific Signal and Information Processing Association Annual Summit and Conference (APSIPA ASC)},
  pages={846--853},
  year={2023},
  organization={IEEE}
}

@inproceedings{zhou2023synthetic,
  title={Synthetic lies: Understanding ai-generated misinformation and evaluating algorithmic and human solutions},
  author={Zhou, Jiawei and Zhang, Yixuan and Luo, Qianni and Parker, Andrea G and De Choudhury, Munmun},
  booktitle={Proceedings of the 2023 CHI conference on human factors in computing systems},
  pages={1--20},
  year={2023}
}

@inproceedings{park2023generative,
  title={Generative agents: Interactive simulacra of human behavior},
  author={Park, Joon Sung and O'Brien, Joseph and Cai, Carrie Jun and Morris, Meredith Ringel and Liang, Percy and Bernstein, Michael S},
  booktitle={Proceedings of the 36th annual acm symposium on user interface software and technology},
  pages={1--22},
  year={2023}
}

@inproceedings{devlin2019bert,
  title={Bert: Pre-training of deep bidirectional transformers for language understanding},
  author={Devlin, Jacob and Chang, Ming-Wei and Lee, Kenton and Toutanova, Kristina},
  booktitle={Proceedings of the 2019 conference of the North American chapter of the association for computational linguistics: human language technologies, volume 1 (long and short papers)},
  pages={4171--4186},
  year={2019}
}

@article{peng2024not,
  title={Not all fake news is semantically similar: Contextual semantic representation learning for multimodal fake news detection},
  author={Peng, Liwen and Jian, Songlei and Kan, Zhigang and Qiao, Linbo and Li, Dongsheng},
  journal={Information Processing \& Management},
  volume={61},
  number={1},
  pages={103564},
  year={2024},
  publisher={Elsevier}
}

@article{shu2020fakenewsnet,
  title={Fakenewsnet: A data repository with news content, social context, and spatiotemporal information for studying fake news on social media},
  author={Shu, Kai and Mahudeswaran, Deepak and Wang, Suhang and Lee, Dongwon and Liu, Huan},
  journal={Big data},
  volume={8},
  number={3},
  pages={171--188},
  year={2020},
  publisher={Mary Ann Liebert, Inc., publishers 140 Huguenot Street, 3rd Floor New~…}
}

@article{zhang2019d,
  title={D-vae: A variational autoencoder for directed acyclic graphs},
  author={Zhang, Muhan and Jiang, Shali and Cui, Zhicheng and Garnett, Roman and Chen, Yixin},
  journal={Advances in neural information processing systems},
  volume={32},
  year={2019}
}

@inproceedings{park2025adversarial,
  title={Adversarial Style Augmentation via Large Language Model for Robust Fake News Detection},
  author={Park, Sungwon and Han, Sungwon and Xie, Xing and Lee, Jae-Gil and Cha, Meeyoung},
  booktitle={Proceedings of the ACM on Web Conference 2025},
  pages={4024--4033},
  year={2025}
}

@inproceedings{liu2025modality,
  title={Modality interactive mixture-of-experts for fake news detection},
  author={Liu, Yifan and Liu, Yaokun and Li, Zelin and Yao, Ruichen and Zhang, Yang and Wang, Dong},
  booktitle={Proceedings of the ACM on Web Conference 2025},
  pages={5139--5150},
  year={2025}
}

@inproceedings{jin2017multimodal,
  title={Multimodal fusion with recurrent neural networks for rumor detection on microblogs},
  author={Jin, Zhiwei and Cao, Juan and Guo, Han and Zhang, Yongdong and Luo, Jiebo},
  booktitle={Proceedings of the 25th ACM international conference on Multimedia},
  pages={795--816},
  year={2017}
}

@article{wang2024survey,
  title={A survey on large language model based autonomous agents},
  author={Wang, Lei and Ma, Chen and Feng, Xueyang and Zhang, Zeyu and Yang, Hao and Zhang, Jingsen and Chen, Zhiyuan and Tang, Jiakai and Chen, Xu and Lin, Yankai and others},
  journal={Frontiers of Computer Science},
  volume={18},
  number={6},
  pages={186345},
  year={2024},
  publisher={Springer}
}

@article{li2024agent,
  title={Agent hospital: A simulacrum of hospital with evolvable medical agents},
  author={Li, Junkai and Lai, Yunghwei and Li, Weitao and Ren, Jingyi and Zhang, Meng and Kang, Xinhui and Wang, Siyu and Li, Peng and Zhang, Ya-Qin and Ma, Weizhi and others},
  journal={arXiv preprint arXiv:2405.02957},
  year={2024}
}

@inproceedings{bakshy2012role,
  title={The role of social networks in information diffusion},
  author={Bakshy, Eytan and Rosenn, Itamar and Marlow, Cameron and Adamic, Lada},
  booktitle={Proceedings of the 21st international conference on World Wide Web},
  pages={519--528},
  year={2012}
}

@article{douze2024faiss,
  title={The faiss library},
  author={Douze, Matthijs and Guzhva, Alexandr and Deng, Chengqi and Johnson, Jeff and Szilvasy, Gergely and Mazar{\'e}, Pierre-Emmanuel and Lomeli, Maria and Hosseini, Lucas and J{\'e}gou, Herv{\'e}},
  journal={arXiv preprint arXiv:2401.08281},
  year={2024}
}

@article{wang2025user,
  title={User behavior simulation with large language model-based agents},
  author={Wang, Lei and Zhang, Jingsen and Yang, Hao and Chen, Zhi-Yuan and Tang, Jiakai and Zhang, Zeyu and Chen, Xu and Lin, Yankai and Sun, Hao and Song, Ruihua and others},
  journal={ACM Transactions on Information Systems},
  volume={43},
  number={2},
  pages={1--37},
  year={2025},
  publisher={ACM New York, NY}
}

@techreport{horton2023large,
  title={Large language models as simulated economic agents: What can we learn from homo silicus?},
  author={Horton, John J},
  year={2023},
  institution={National Bureau of Economic Research}
}

@article{xie2024can,
  title={Can large language model agents simulate human trust behavior?},
  author={Xie, Chengxing and Chen, Canyu and Jia, Feiran and Ye, Ziyu and Lai, Shiyang and Shu, Kai and Gu, Jindong and Bibi, Adel and Hu, Ziniu and Jurgens, David and others},
  journal={Advances in neural information processing systems},
  volume={37},
  pages={15674--15729},
  year={2024}
}

@article{xu2023exploring,
  title={Exploring large language models for communication games: An empirical study on werewolf},
  author={Xu, Yuzhuang and Wang, Shuo and Li, Peng and Luo, Fuwen and Wang, Xiaolong and Liu, Weidong and Liu, Yang},
  journal={arXiv preprint arXiv:2309.04658},
  year={2023}
}

@article{light2023avalonbench,
  title={Avalonbench: Evaluating llms playing the game of avalon},
  author={Light, Jonathan and Cai, Min and Shen, Sheng and Hu, Ziniu},
  journal={arXiv preprint arXiv:2310.05036},
  year={2023}
}

@article{lewis2020retrieval,
  title={Retrieval-augmented generation for knowledge-intensive nlp tasks},
  author={Lewis, Patrick and Perez, Ethan and Piktus, Aleksandra and Petroni, Fabio and Karpukhin, Vladimir and Goyal, Naman and K{\"u}ttler, Heinrich and Lewis, Mike and Yih, Wen-tau and Rockt{\"a}schel, Tim and others},
  journal={Advances in neural information processing systems},
  volume={33},
  pages={9459--9474},
  year={2020}
}

@inproceedings{hu2025decomposition,
  title={Decomposition Dilemmas: Does Claim Decomposition Boost or Burden Fact-Checking Performance?},
  author={Hu, Qisheng and Long, Quanyu and Wang, Wenya},
  booktitle={Proceedings of the 2025 Conference of the Nations of the Americas Chapter of the Association for Computational Linguistics: Human Language Technologies (Volume 1: Long Papers)},
  pages={6313--6336},
  year={2025}
}

@inproceedings{li2024self,
  title={Self-checker: Plug-and-play modules for fact-checking with large language models},
  author={Li, Miaoran and Peng, Baolin and Galley, Michel and Gao, Jianfeng and Zhang, Zhu},
  booktitle={Findings of the Association for Computational Linguistics: NAACL 2024},
  pages={163--181},
  year={2024}
}

@inproceedings{ma2025local,
  title={Local: Logical and causal fact-checking with llm-based multi-agents},
  author={Ma, Jiatong and Hu, Linmei and Li, Rang and Fu, Wenbo},
  booktitle={Proceedings of the ACM on Web Conference 2025},
  pages={1614--1625},
  year={2025}
}

@article{han2025debate,
  title={Debate-to-Detect: Reformulating Misinformation Detection as a Real-World Debate with Large Language Models},
  author={Han, Chen and Zheng, Wenzhen and Tang, Xijin},
  journal={arXiv preprint arXiv:2505.18596},
  year={2025}
}

@article{velivckovic2017graph,
  title={Graph attention networks},
  author={Veli{\v{c}}kovi{\'c}, Petar and Cucurull, Guillem and Casanova, Arantxa and Romero, Adriana and Lio, Pietro and Bengio, Yoshua},
  journal={arXiv preprint arXiv:1710.10903},
  year={2017}
}

@article{liu2024deepseek,
  title={Deepseek-v3 technical report},
  author={Liu, Aixin and Feng, Bei and Xue, Bing and Wang, Bingxuan and Wu, Bochao and Lu, Chengda and Zhao, Chenggang and Deng, Chengqi and Zhang, Chenyu and Ruan, Chong and others},
  journal={arXiv preprint arXiv:2412.19437},
  year={2024}
}

@inproceedings{wang2025collaboration,
  title={Collaboration and Controversy Among Experts: Rumor Early Detection by Tuning a Comment Generator},
  author={Wang, Bing and Zhao, Bingrui and Li, Ximing and Li, Changchun and Gao, Wanfu and Wang, Shengsheng},
  booktitle={Proceedings of the 48th International ACM SIGIR Conference on Research and Development in Information Retrieval},
  pages={468--478},
  year={2025}
}

@inproceedings{li2025multi,
  title={A Multi-Agent Framework with Automated Decision Rule Optimization for Cross-Domain Misinformation Detection},
  author={Li, Hui and Wang, Ante and Li, Kunquan and Wang, Zhihao and Zhang, Liang and Qiu, Delai and Liu, Qingsong and Su, Jinsong},
  booktitle={Proceedings of the 2025 Conference on Empirical Methods in Natural Language Processing},
  pages={5720--5736},
  year={2025}
}

@article{liu2020fned,
  title={Fned: a deep network for fake news early detection on social media},
  author={Liu, Yang and Wu, Yi-Fang Brook},
  journal={ACM Transactions on Information Systems (TOIS)},
  volume={38},
  number={3},
  pages={1--33},
  year={2020},
  publisher={ACM New York, NY, USA}
}

@article{zhang2020adversarial,
  title={Adversarial attacks on deep-learning models in natural language processing: A survey},
  author={Zhang, Wei Emma and Sheng, Quan Z and Alhazmi, Ahoud and Li, Chenliang},
  journal={ACM Transactions on Intelligent Systems and Technology (TIST)},
  volume={11},
  number={3},
  pages={1--41},
  year={2020},
  publisher={ACM New York, NY, USA}
}

@article{shu2017fake,
  title={Fake news detection on social media: A data mining perspective},
  author={Shu, Kai and Sliva, Amy and Wang, Suhang and Tang, Jiliang and Liu, Huan},
  journal={ACM SIGKDD explorations newsletter},
  volume={19},
  number={1},
  pages={22--36},
  year={2017},
  publisher={ACM New York, NY, USA}
}

@article{liu2024tiny,
  title={From a tiny slip to a giant leap: An llm-based simulation for fake news evolution},
  author={Liu, Yuhan and Song, Zirui and Zhang, Xiaoqing and Chen, Xiuying and Yan, Rui},
  journal={arXiv e-prints},
  pages={arXiv--2410},
  year={2024}
}

@inproceedings{hu2024llm,
  title={An LLM-enhanced agent-based simulation tool for information propagation},
  author={Hu, Yuxuan and Sherpa, Gemju and Zhang, Lan and Li, Weihua and Bai, Quan and Wang, Yijun and Wang, Xiaodan},
  booktitle={Proceedings of the Thirty-Third International Joint Conference on Artificial Intelligence, IJCAI-24, Jeju, Republic of Korea},
  pages={3--9},
  year={2024}
}

@article{hu2025simulating,
  title={Simulating rumor spreading in social networks using llm agents},
  author={Hu, Tianrui and Liakopoulos, Dimitrios and Wei, Xiwen and Marculescu, Radu and Yadwadkar, Neeraja J},
  journal={arXiv preprint arXiv:2502.01450},
  year={2025}
}

@article{wang2025yulan,
  title={Yulan-onesim: Towards the next generation of social simulator with large language models},
  author={Wang, Lei and Gao, Heyang and Bo, Xiaohe and Chen, Xu and Wen, Ji-Rong},
  journal={arXiv preprint arXiv:2505.07581},
  year={2025}
}

@inproceedings{ju2025trajllm,
  title={Trajllm: A modular llm-enhanced agent-based framework for realistic human trajectory simulation},
  author={Ju, Chenlu and Liu, Jiaxin and Sinha, Shobhit and Xue, Hao and Salim, Flora},
  booktitle={Companion Proceedings of the ACM on Web Conference 2025},
  pages={2847--2850},
  year={2025}
}

@inproceedings{dou2021user,
  title={User preference-aware fake news detection},
  author={Dou, Yingtong and Shu, Kai and Xia, Congying and Yu, Philip S and Sun, Lichao},
  booktitle={Proceedings of the 44th international ACM SIGIR conference on research and development in information retrieval},
  pages={2051--2055},
  year={2021}
}

@article{wei2022chain,
  title={Chain-of-thought prompting elicits reasoning in large language models},
  author={Wei, Jason and Wang, Xuezhi and Schuurmans, Dale and Bosma, Maarten and Xia, Fei and Chi, Ed and Le, Quoc V and Zhou, Denny and others},
  journal={Advances in neural information processing systems},
  volume={35},
  pages={24824--24837},
  year={2022}
}

@misc{kingma2013auto,
  title={Auto-encoding variational bayes},
  author={Kingma, Diederik P and Welling, Max and others},
  year={2013},
  publisher={Banff, Canada}
}

@article{kipf2016semi,
  title={Semi-Supervised Classification with Graph Convolutional Networks},
  author={Kipf, TN},
  journal={arXiv preprint arXiv:1609.02907},
  year={2016}
}

@inproceedings{sun2022rumor,
  title={Rumor detection on social media with graph adversarial contrastive learning},
  author={Sun, Tiening and Qian, Zhong and Dong, Sujun and Li, Peifeng and Zhu, Qiaoming},
  booktitle={Proceedings of the ACM web conference 2022},
  pages={2789--2797},
  year={2022}
}

@inproceedings{you2018graphrnn,
  title={Graphrnn: Generating realistic graphs with deep auto-regressive models},
  author={You, Jiaxuan and Ying, Rex and Ren, Xiang and Hamilton, William and Leskovec, Jure},
  booktitle={International conference on machine learning},
  pages={5708--5717},
  year={2018},
  organization={PMLR}
}

@inproceedings{zhang2024mitigating,
  title={Mitigating social hazards: Early detection of fake news via diffusion-guided propagation path generation},
  author={Zhang, Litian and Zhang, Xiaoming and Li, Chaozhuo and Zhou, Ziyi and Liu, Jiacheng and Huang, Feiran and Zhang, Xi},
  booktitle={Proceedings of the 32nd ACM International Conference on Multimedia},
  pages={2842--2851},
  year={2024}
}

@inproceedings{bian2020rumor,
  title={Rumor detection on social media with bi-directional graph convolutional networks},
  author={Bian, Tian and Xiao, Xi and Xu, Tingyang and Zhao, Peilin and Huang, Wenbing and Rong, Yu and Huang, Junzhou},
  booktitle={Proceedings of the AAAI conference on artificial intelligence},
  volume={34},
  number={01},
  pages={549--556},
  year={2020}
}

@article{yao2007early,
  title={On early stopping in gradient descent learning},
  author={Yao, Yuan and Rosasco, Lorenzo and Caponnetto, Andrea},
  journal={Constructive approximation},
  volume={26},
  number={2},
  pages={289--315},
  year={2007},
  publisher={Springer}
}

@article{kingma2014adam,
  title={Adam: A method for stochastic optimization},
  author={Kingma, Diederik P},
  journal={arXiv preprint arXiv:1412.6980},
  year={2014}
}

@inproceedings{zhu2024propagation,
  title={Propagation structure-aware graph transformer for robust and interpretable fake news detection},
  author={Zhu, Junyou and Gao, Chao and Yin, Ze and Li, Xianghua and Kurths, J{\"u}rgen},
  booktitle={Proceedings of the 30th ACM SIGKDD Conference on Knowledge Discovery and Data Mining},
  pages={4652--4663},
  year={2024}
}

@inproceedings{liu2025mosaic,
  title={Mosaic: Modeling social ai for content dissemination and regulation in multi-agent simulations},
  author={Liu, Genglin and Le, Vivian T and Rahman, Salman and Kreiss, Elisa and Ghassemi, Marzyeh and Gabriel, Saadia},
  booktitle={Proceedings of the 2025 Conference on Empirical Methods in Natural Language Processing},
  pages={6401--6428},
  year={2025}
}

@article{kiesling2012agent,
  title={Agent-based simulation of innovation diffusion: a review},
  author={Kiesling, Elmar and G{\"u}nther, Markus and Stummer, Christian and Wakolbinger, Lea M},
  journal={Central European Journal of Operations Research},
  volume={20},
  number={2},
  pages={183--230},
  year={2012},
  publisher={Springer}
}

@article{kaliyar2021fakebert,
  title={FakeBERT: Fake news detection in social media with a BERT-based deep learning approach},
  author={Kaliyar, Rohit Kumar and Goswami, Anurag and Narang, Pratik},
  journal={Multimedia tools and applications},
  volume={80},
  number={8},
  pages={11765--11788},
  year={2021},
  publisher={Springer}
}

@article{pavlyshenko2023analysis,
  title={Analysis of disinformation and fake news detection using fine-tuned large language model},
  author={Pavlyshenko, Bohdan M},
  journal={arXiv preprint arXiv:2309.04704},
  year={2023}
}

@inproceedings{hou2024dag,
  title={Dag-aware variational autoencoder for social propagation graph generation},
  author={Hou, Dongpeng and Gao, Chao and Li, Xuelong and Wang, Zhen},
  booktitle={Proceedings of the AAAI Conference on Artificial Intelligence},
  volume={38},
  number={8},
  pages={8508--8516},
  year={2024}
}

@inproceedings{ding2020bert,
  title={BERT-based mental model, a better fake news detector},
  author={Ding, Jia and Hu, Yongjun and Chang, Huiyou},
  booktitle={Proceedings of the 2020 6th international conference on computing and artificial intelligence},
  pages={396--400},
  year={2020}
}

@article{shi2025personax,
  title={Personax: A recommendation agent oriented user modeling framework for long behavior sequence},
  author={Shi, Yunxiao and Xu, Wujiang and Zhang, Zeqi and Zi, Xing and Wu, Qiang and Xu, Min},
  journal={arXiv preprint arXiv:2503.02398},
  year={2025}
}

@article{zhang2019empirically,
  title={Empirically grounded agent-based models of innovation diffusion: a critical review},
  author={Zhang, Haifeng and Vorobeychik, Yevgeniy},
  journal={Artificial Intelligence Review},
  volume={52},
  number={1},
  pages={707--741},
  year={2019},
  publisher={Springer}
}

@inproceedings{zheng2022mfan,
  title={MFAN: Multi-modal Feature-enhanced Attention Networks for Rumor Detection.},
  author={Zheng, Jiaqi and Zhang, Xi and Guo, Sanchuan and Wang, Quan and Zang, Wenyu and Zhang, Yongdong},
  booktitle={IJCAI},
  volume={2022},
  pages={2413--2419},
  year={2022}
}

@inproceedings{nan2024let,
  title={Let silence speak: Enhancing fake news detection with generated comments from large language models},
  author={Nan, Qiong and Sheng, Qiang and Cao, Juan and Hu, Beizhe and Wang, Danding and Li, Jintao},
  booktitle={Proceedings of the 33rd ACM International Conference on Information and Knowledge Management},
  pages={1732--1742},
  year={2024}
}










\end{document}